\documentclass[useAMS,usenatbib]{mnras}
\usepackage[english]{babel}
\usepackage{graphicx}

\title[Cosmological filaments]{A comparison of cosmological filaments catalogues}
\author[A. Rost et al.]{Agust\'in Rost,$^{1,2}$\thanks{E-mail: $a\_rost@unc.edu.ar$} 
Federico Stasyszyn,$^{1,2}$ Luis Pereyra$^{1,2}$ 
and H\'ector J. Mart\'inez$^{1,2}$ \\
\\
$^{1}$ Instituto de Astronom\'ia Te\'orica y Experimental (IATE), CONICET - UNC, Laprida 854,
X5000BGR, C\'ordoba, Argentina\\
$^{2}$ Observatorio Astron\'omico, Universidad Nacional de C\'ordoba, Laprida 854, X5000BGR, C\'ordoba, Argentina.\\
}

\begin{document}

\date{Accepted ???. Received ???; in original form ???}

\maketitle

\label{firstpage}

\begin{abstract}

In this work we compare three catalogues of cosmological filaments identified in
the Sloan Digital Sky Survey by means of different algorithms by Tempel et al., Pereyra
et al., and Mart\'inez et al.
We analyse how different identification techniques determine differences
in the filament statistical properties: length, elongation, redshift distribution,
and abundance.
We find that the statistical properties of the filaments strongly depend on the 
identification algorithm.
We use a volume limited sample of galaxies to characterise other properties of filaments
such as: galaxy overdensity, luminosity function of galaxies, mean galaxy luminosity, filament
luminosity, and the overdensity profile of galaxies around filaments.
In general, we find that these properties primarily depended on filament length. 
Shorter filaments have larger overdensities, are more populated by red galaxies, 
and have better defined galaxy overdensity profiles, than longer filaments. 
Concluding that galaxies belonging to filaments have 
characteristic signatures depending on the identification  algorithm used.

\end{abstract}


\section{Introduction} 

Structures in the Universe are organised hierarchically. From the large scales
that are not even bound to small scales as galaxies that act as the
cosmological building blocks, all the systems are related. The distribution of
matter is a complex web-like structure that includes galaxy clusters, super-clusters,
voids, walls and filaments. The filaments are long shaped structures of matter
with intermediate densities, galaxy clusters are round shaped with higher
densities (above a critical density $\rho_{\rm crit}$), walls are planar structures, 
more extended than filaments and with overall lower densities. These structures are entangled,
they connect the high density peaks, and leave the majority of the
volume empty with the so called voids \citep{bond} (with typically densities
under $0.8 \rho_{\rm crit}$), which tend to become rounder as they evolve.  
It is not trivial to define precisely, and unequivocally, any of 
these structures. This is due to their intrinsic complexity, wide range of
densities, and diffuse barriers between them. The definition of these structures 
depends strongly on whether they are to be identified in simulations or in observations. 
One way to characterise the differences between these structures is to consider clusters,
filaments, and walls, as regions with maximal matter density, but with different
number of positive eigenvalues of the Hessian matrix of the density field. For
the case of clusters, the eigenvalues are all negative, for walls two of
them are positive, and for filaments just one  \citep[e.g.][]{chen}.

There have been different ways to define filaments in the literature.
According to \citet{pogosyan1} these objects can be thought as matter ridges that have 
high densities at the extremes, and a saddle point near the centre. They could also have
substructures and then high density peaks near the centre, as discussed by \citet{cautun}. 
Some algorithms to detect filaments take this definition into account. Other authors
define filaments as structures that connect massive halos and are
traced by other, lesser massive, halos, disregarding the existence of a saddle point
between them \citep[e.g.][]{Alpaslan2014, Park2009}. 
There are works that are based on visual inspection
\citep{suzaku}, identification of regions inter-clusters \citep{Pimbblet2004,
Colberg2005}, or on a random configuration of galaxies based on a
modelled probability distribution \citep{tempel}. There are even
post-processing algorithms that look for iso-densities surfaces of gas in simulations 
\citep{gheller}.

Another example in the literature is {\tt DisPerSE} \citep{disperse} 
a scale-free and parameter-free method to detect nodes, filaments and walls.
{\tt DisPerSE} identifies structures as components of 
the Morse-Smale complex of an input density field calculated with
the Delaunay tessellation field estimator (DTFE) \citep{Schaap2000}. 
This method proved that it is flexible and robust with both cosmological simulations and observational catalogues. For example \citet{Kleiner} use this algorithm to find filaments in the 6dF \citep{6dF1, 6dF2} covering the entire southern sky and \citet{kraljic} in GAMA survey \citep{gamma1, gamma2, gamma3, gamma4}.
It should be noted that not all algorithms can identify walls explicitly, those methods that do not 
detect walls may encounter difficulties when dealing with these objects, for instance, 
by disarming walls as different filaments. 

Despite the various drawbacks that the identification of cosmological filaments possess,
the study and characterisation of filaments are interesting for different reasons.
Several works suggest that in
these regions there are unique astrophysical phenomena taking place, corroborating
the idea that their nature is related but independent of other structures
\citep{smith}.

\citet{Tempel2013} investigate the alignment of spiral/elliptical galaxies, they find that the minor axis
of elliptical galaxies tends to be preferentially perpendicular to the hosting filament's axis, meanwhile 
bright spiral galaxies tend to be aligned with the host filament's axis. 
\citet{Zhang2015} study the alignment between the spin axis of spiral galaxies and the filament 
direction. They show that the spin axis of spiral galaxies in filaments tends to be preferentially 
perpendicular to the direction of filaments. The lack of consensus on the observation side regarding 
galaxy-filament alignments may be due, among other effects, to differences in the definition and method of identification, as well as to unwanted artefacts in sample selection.

\citet{codis} describe the velocity field around filamentary structures,
proposing that this velocity field can be explained with Zeldovich
approximation at the saddle points, relating them with the orientation and
shapes of galaxies in these environments. 
\citet{kraljic} study filaments as {\it highways} of galaxies that are important in their evolution, 
this work also studies walls as unique objects. 
Another example is \citet{martinez} where they find that galaxies in filaments linking groups
of galaxies have lower specific star formation rates, and therefore are more quenched, than galaxies 
that are infalling into groups from other directions.
A similar result is found by \citet{Salerno2019} at higher redshifts. This work provides evidence of
a distinctive environmental effect by filaments upon galaxies as early as $z\sim0.9$.
\citet{libeskind} reviews and compares different approaches to define filaments, stressed on their detection in numerical simulations. 

In this paper, we focus in algorithms for detecting filaments, applied over the same sample
of galaxies from the Sloan Digital Sky Survey (SDSS, \citealt{York:2000}).
We compare three filaments samples found with different methods, showing
the properties and characteristics of each of them.  Although it would have
been better to add more catalogues to the comparison, it was not possible for
some of them, the majority of the algorithms studied by \citet{cautun},  work
with simulations. The methods are much more limited for observational
catalogues due to the lack of information in general, for example the soft
distribution of dark matter, the algorithms also have to deal with the
projected velocity dispersion (the effect {\it Fingers of God}) and the typical
incompleteness of galaxy catalogues. However, some of the algorithms that work
with observations could not be studied, \citet{chen} catalogue for example
requires high computation power to consider a reasonable region from the SDSS
catalogue with no restriction in redshift.

This paper is organised as follows: in section \ref{sec:data} we present the galaxy and filament
samples used in this work, in section \ref{sec:GralProp} we show general properties of the 
filament catalogues we use and, in section \ref{sec:analysis} we develop the tools that will be used on
galaxy catalogues and find statistical properties of filaments and finally we write our conclusions in
in section \ref{sec:conclusions}.

\section{Observational samples} \label{sec:data}

In this paper we analyse three filament catalogues built using different types of
methods and observational samples. 
In this section we describe each of them to later on study their
differences.  We call {\it nodes} to the filament extreme points, while we
refer to the points that form a filamentary path as {\it path nodes}.

\subsection{Minimal spanning tree (P19)}

The sample of filaments by \cite{pereyra_submitted} is constructed by means of a cosmological filament finding algorithm based on
a technique borrowed from graph theory: the minimal spanning tree ({\it
MST}). The authors adopt the definitions of \citet{MST} and also previously 
used by others \citep[e.g.][]{Colberg2007,Alpaslan2014}.
A graph is a set of vertex (centres of dark matter halos for numerical
simulations, or galaxies for observations typically), edges (they connect vertices)
and weights.
A {\it MST} is the unique set of edges (if all the weights are different) 
that efficiently connects all the vertices from the initial graph, without closed cycles, and
resulting in a minimal sum of the weighs.

The {\it MST} describes mainly the distribution of close neighbours and eventually, 
is capable of making a diffuse {\it MST} when the number of vertices is too high, and 
therefore unable to characterise properly the large scale structure \citep{Stoica2005}.
To prevent these drawbacks, the authors limit the {\it MST} to a intermediate density region. 
This region is identified by a {\it Friends-of-Friends} (FoF) algorithm 
designed specifically for flux-limited galaxy surveys, as described in \citet{Merzan2005}. 
\citet{pereyra_submitted} use a transverse large link length of $1.24 \, h^{-1} {\rm Mpc}$ corresponding 
to an overdensity of $\delta\rho / \rho = 1$ and a line-of-sight link length of $V_{0} = 200\ 
\mathrm{km\ s^{-1}}$. In addition, each edge in the graph is weighted by the luminosity of the 
galaxies at its ends.

By using data from SDSS DR12 \citep{SLOAN_DR12}, 
they build the {\it MST}, with all the bright galaxies ($M_{r} < -20.5$) 
included in the intermediate density region as path nodes. 
Then they extract the different branches of the {\it MST}. 
\citet{pereyra_submitted} consider as filaments those branches of the tree that have galaxies brighter 
than $ m_ {r} <-21.0 $ at their extremes. As a result, a set of $47249$ filaments is achieved.
The final catalogue contains the physical length of the filament, elongation, RMS, 
number of galaxies that conform the spine of filament 
and IDs of the galaxies in it, as well as the spectroscopic and photometric properties of the 
galaxies which were used to build the filament sample. 

Hereafter we will refer to this sample of filaments as P19.

\subsection{Bisous model (T14)}

\citet{tempel} use an algorithm called {\it Bisous model} 
which approximates the filamentary structure as
multiple cylinders along the filaments that indicate the probability density of
galaxies, and the galaxy distribution as a random sample following this
density. To determine this structure of cylinders, they start with
a random configuration of cylinders that, step by step, is advanced with a
stochastic process based on Markov-chains Monte Carlo.
The algorithm advances the cylinder locations and orientations to fit with the galaxy distribution. 
One of the advantages of this approach is that the process 
relies on the galaxy positions, and no pre-processing, such as computing a smooth density field, 
is needed. This method is purely based on the geometrical distribution of galaxies, 
it does not use information about their luminosity or mass.

The filament catalogue is built using the distribution
of galaxies in the spectroscopic galaxy sample of SDSS DR8 \citep{SLOAN_DR8}. 
The catalogue is obtained after concatenating straight lines that have $0.5 {\rm~ Mpc}$ of length. 
Resulting node's positions do not necessarily coincide with galaxies or galaxy cluster positions.
The catalogue contains the following information about filaments:
length, total luminosity of the filament, number of path nodes, co-moving coordinates of each point.
There is complementary information for the galaxies associated to these filaments:
id of the nearest filament, id of the nearest filament point, and
distance from the nearest filament spine, to name only a few.
For a detailed description of the filament catalogue see \citep{tempel}.

Hereafter we will refer to this sample of filaments as T14.

\subsection{Filaments linking groups of galaxies (M16)}

\citet{martinez} build a sample of filaments using a galaxy group catalogue from
\citet{ZandivarezMartinez2011} as nodes.
Filaments consist of pairs of close galaxy groups that are linked by an overdensity 
of galaxies (see their Fig. 1).
Curved or branched structures are not considered because, by definition, these filaments are 
straight lines joining the nodes.  Only groups with virial masses above the average of the 
group catalogue $\log(M_{\rm vir}/h^{-1} M_{\odot}) > 13.5$, and in the redshift range 
$0.05\le z\le 0.15$ are used.  
The authors consider that a pair of group is linked by a filament if:
(i) they are separated by less than $1000~ {\rm km ~s^{-1}}$ in radial velocity;
(ii) their projected distance is less than $10 h^{-1}~ {\rm Mpc}$;
(iii) the galaxy overdensity in a cuboid-like region between the nodes that is centred on 
the barycentre of the pair, is greater than a threshold (see details in \citealt{martinez}).

As result of the identification, at a given filament, 
we find the physical properties of the nodes such as
separation between groups, number of galaxies members, 
virial radius, virial mass, velocity dispersion.
This filament sample has been constructed to study the environmental effects of filaments upon 
galaxies that are infalling into groups and is not intended to be complete.

Hereafter we will refer to this sample of filaments as M16.

\subsection{Galaxy sample}

For a fair comparison between the three filament catalogues, we use the same parent galaxy 
catalogue and search for galaxies in this catalogue that lie in the filaments. 
We use the catalogue of galaxies by \citet{Tempel2017yCat}, which was downloaded from the
SDSS catalogue Archive Server
\citep[CAS\footnote{\url{http://skyserver.sdss3.org/casjobs/}},][]{Eisenstein2011, SLOAN_DR12}. These authors added
redshifts originated from the Two-degree Field Galaxy Redshift Survey \citep{2dFGRS, Colless2003},
the Two Micron All Sky Survey Redshift Survey \citep{2MRS}, and the Third Reference
Catalogue of Bright Galaxies \citep[RC3,][]{Vaucouleurs1991,Corwin1994}. See \citet{Tempel2014} for more details.
We have adopted the {\tt ModelMag} magnitudes corrected by extinction and then
applied the offset and the $k-$correction following the empirical $k-$correction of
\citet{omill2011} at $z = 0.1$.  In order to minimise the inclusion of
foreground stars \citep{collister2007} we used only galaxies with $(g - r) < 3$
mag.

The volume-complete set of galaxies is determined setting an upper limit of $z
\leq 0.137$ and a maximum value of absolute magnitude $M_r=-20.43$, which is computed
assuming $H_{0} = 100~ {\rm km~s^{-1}~Mpc^{-1}}$, $\Omega_{m} = 0.3$ and
$\Omega_{\Lambda} = 0.7$ cosmology. We are aware of possible star-contamination and fiber collisions of 
the sample, however we find treatment done by \citet{Tempel2017yCat}, and \citet{omill2011}, good enough for our 
purposes. This is the catalogue of tracers that we use in this work.

\subsection{Random sample of galaxies}

We use a random catalogue of galaxies distributed over the same angular
distribution of our galaxy catalogue. 
It consists of $\sim 30,000,000$ galaxies with
$(\alpha, \delta)$ coordinates and redshift, $45$ times denser than the real
galaxy sample.  The random sample is a cloning procedure in which every galaxy in our volume
limited sample of galaxies is cloned 45 times by assigning to it a random redshift and angular 
position, bound to mimic the distributions of redshift and angular coverage of the
galaxy sample.
For this purpose, we constructed an angular coverage mask of the SDSS DR12 using 
routines from the software HEALPix\footnote{\url{http://healpix.sourceforge.net}} package \citep{Gorski:2005}.
This procedure does not induce redshift-colour correlations. The relation redshift
vs. magnitude is exactly the same for both real and random catalogues. 

\section{General properties of Filaments}
\label{sec:GralProp}

\begin{figure} 
\centering
\includegraphics[width=0.5\textwidth]{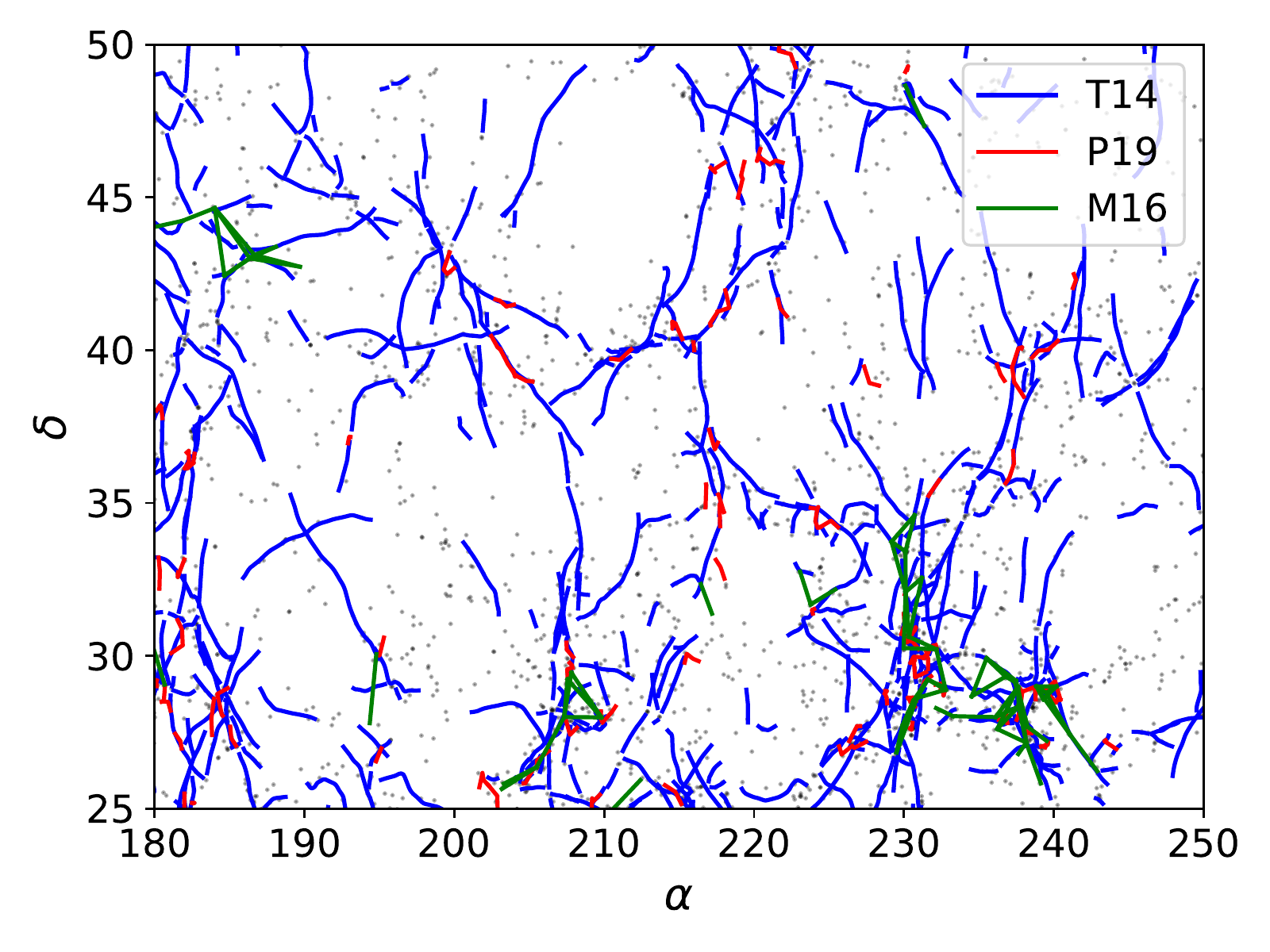} 
\caption[Filaments.] {The three filament samples in the range $z =
0.08 \pm 0.005$, in blue: T14, in red: P19 and in green M16.} 
\label{filaments} 
\end{figure}

\begin{figure} 
\centering
\includegraphics[width=0.5\textwidth]{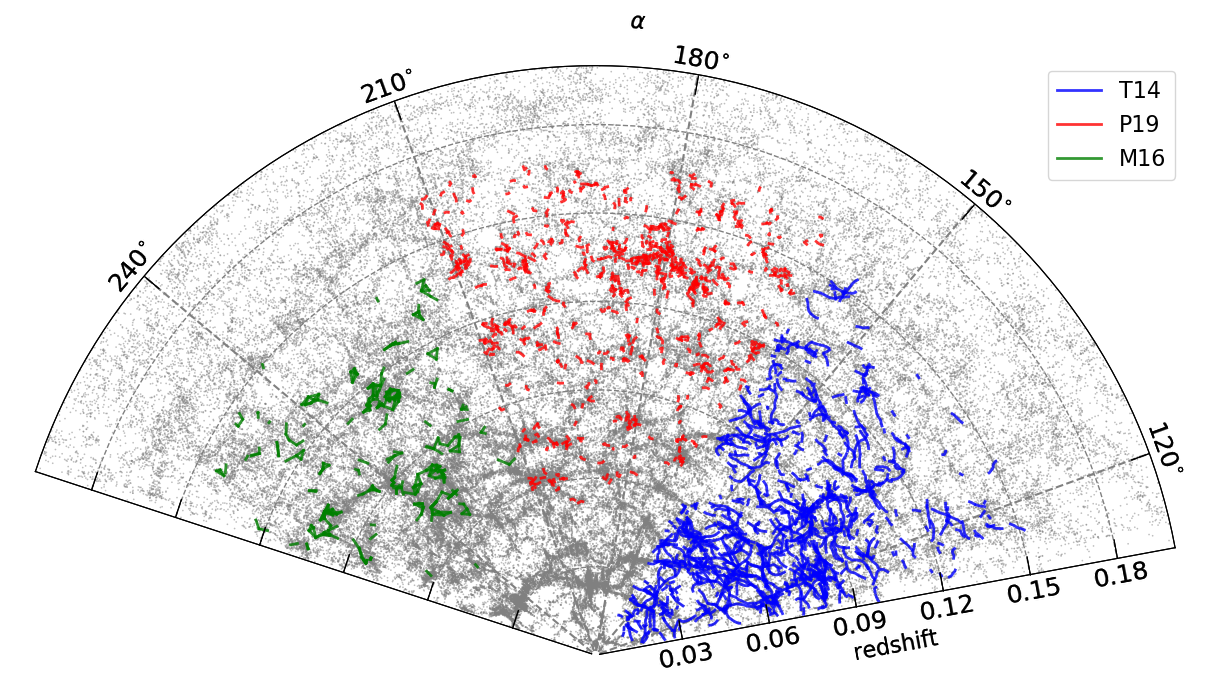} 
\caption[Filaments.] {
The three filament samples in the range $\delta = 32.5 \pm 7.5$, in blue: T14
catalogue, in red: P19 and in green: M16.
}
\label{filaments2} 
\end{figure}

As thoroughly explained above, the three filament catalogues were obtained through 
very different processes.
The T14 catalogue is a sample of $15420$ filaments, while P19 comprises
$8350$, and M16 $3094$. In the Figs. \ref{filaments} and
\ref{filaments2} we show the filaments from the three catalogues over-plotted, in a
redshift slice of $z = 0.005 \pm 0.08$, in the plane of the sky Fig.
\ref{filaments}, and in a slice of $\delta = 25^{\circ} \pm 5^{\circ}$ in Fig.
\ref{filaments2} as a pie in the plane of sight. It can be observed that not
all the filaments are present in the catalogues and they are not evenly found in redshift.
However, there are regions close to bigger structures that seems to cluster the filaments.  

\begin{figure} 
\centering
\includegraphics[width=0.5\textwidth]{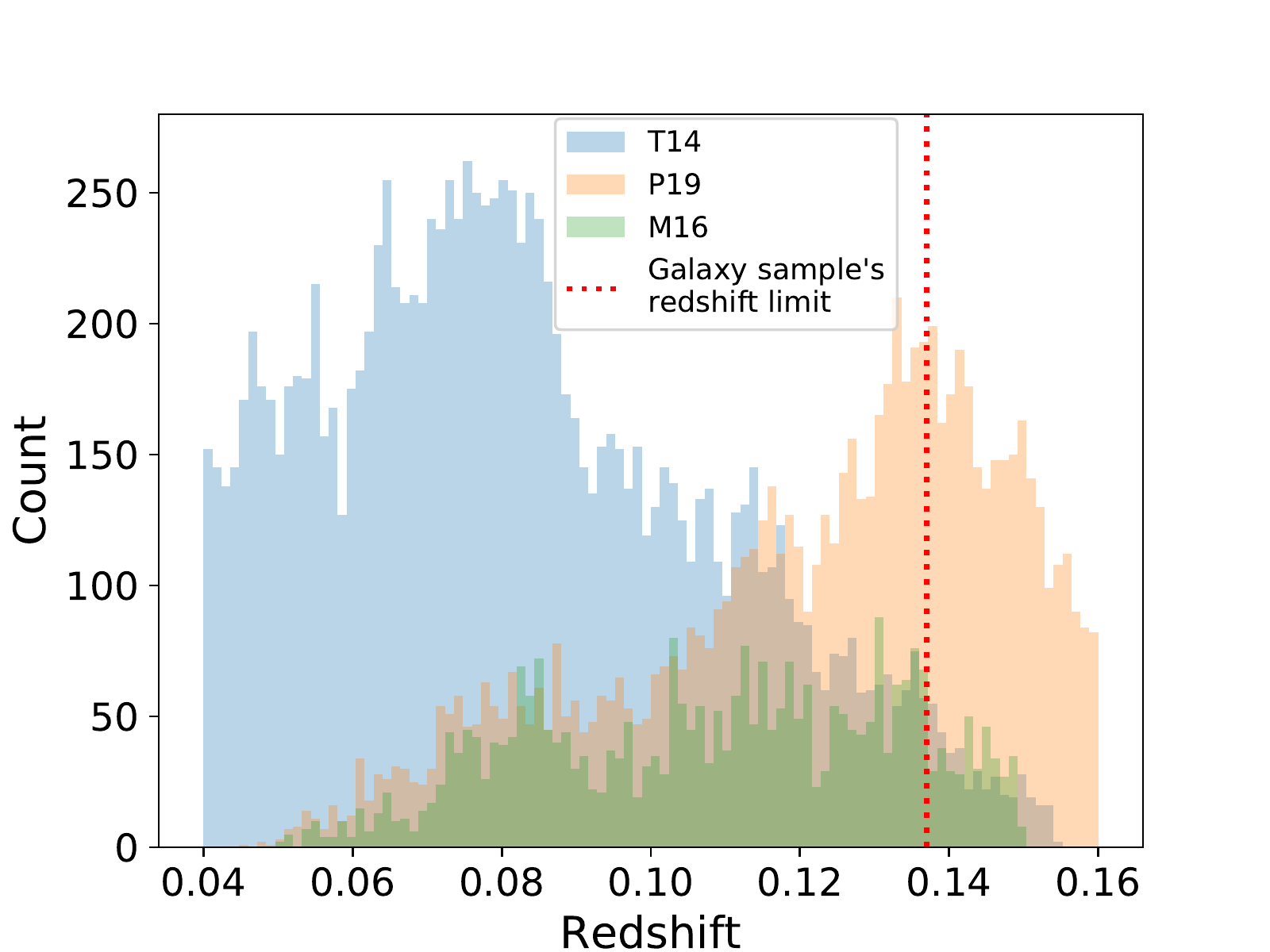} 
\caption[Filaments along z.] {The redshift distribution of the catalogues. 
} 
\label{redshift_distribution} 
\end{figure}

The redshift distribution of the three catalogues is shown in Fig.
\ref{redshift_distribution}, where we can notice that T14 filaments are
substantially closer than P19 and M16, as the peaks are located at
$z = 0.08$, $0.14$ and $0.12$ respectively.  This shows that T14 is
in agreement with the distribution of structures found in \citet{smith}. The
possible relation between properties like the length of filament as a function
of the redshift (as an algorithm bias) has been explored for the catalogues
and we do not find appreciable dependence.  However, the percentage of
galaxies redder than $g - r = 0.7$ slightly increases when decreasing redshift,
with the consequence of higher red fractions in general for T14 filaments.

In what follows, we focus on the projected properties of filaments, therefore, we define a sample 
set of filaments that are perpendicular to the line of sight, and thus in the proper conditions 
to be stacked. To do so, we use the vector
that matches both extremes ${\bf v}_{0, N - 1} = {\bf r}_{N - 1} - {\bf r}_{0}$, where $N$ is the
number of path nodes, and ${\bf r}_{0}$ and ${\bf r}_{N - 1}$ are the position vectors of the
extremes. Then we calculate the cosine of the angle between ${\bf v}_{0, N - 1}$ and ${\bf r}_{0}$: 

\begin{equation} 
\cos(\theta) = \frac{{\bf v}_{0,N - 1} \cdot {\bf r}_{0}}{\vert {\bf v}_{0,N-1} \vert
~ \vert \bf{r}_{0} \vert} 
\end{equation}

If $90^{\circ} - \Delta \theta_{1} <\theta < 90^{\circ} + \Delta \theta_{1}$, 
we consider that the filament belongs to the set $1$ (perpendicular to the view axis), and if 
$\theta < \Delta \theta_{2}$ or $180^{\circ} - \Delta \theta_{2}$ the
filament is in the set $2$ (parallel to the view axis), where $\Delta \theta_{1}$ and $\Delta 
\theta_{2}$ are tolerance angles that can be tuned to increase the number of selected filaments 
or to refine the sample.

In an isotropic and homogeneous Universe, all the directions of the filaments are equally likely,
however observational effects such as the {\it Fingers of God} 
\citep{Jackson1972} determine that the number of detected filaments along the visual axis is usually 
lesser than expected, in particular M16 catalogue excludes this configuration explicitly.

To enhance the signal of the stacking (the proper method is explained in section
\ref{sec:stacking}), we require that the filaments have the same shape through the following parameters:  

\begin{itemize}
\item {\it RMS}: This parameter measures how further away the path nodes are from the positions
that make a straight line between the extremes.
Since ${\bf v}_{0, N - 1}$ is the vector from the start to the end
of the filament, the root mean square of the position of the path nodes from the
straight line is 
\begin{equation}
\textit{RMS} = \sqrt{\frac{1}{N} \sum_{i = 0}^{N - 1} \vert {\bf r}_{i} - {\bf r}_{0}
- {\bf v}_{0,N-1}({\bf r}_{i} - {\bf r}_{0})\cdot {\bf v}_{0,N-1} \vert ^{2} },
\end{equation}
where N is the number of path nodes.

Note that different sized filaments with the same shape will have different
values of {\it RMS}, to solve this, we normalise them by their length to have a scale-independent parameter. High values indicate that the filament is curved or distorted, and the closer the value is to zero, the more similar it is to a straight line.

\item Elongation: This parameter is used by \citet{pereyra_submitted} with their filaments, 
and it is the ratio between the length of the straight line from the extreme and the total length
of the path:

\begin{equation}
\textit{Elong} = \frac{\vert {\bf v}_{0,N-1} \vert}{\sum_{i = 0}^{N - 2} \vert {\bf r}_{i + 1} - 
{\bf r}_{i} \vert}.
\end{equation}

By construction, this value is always equal or less than $1$, the closer the value to $1$,
the more similar the path is to a straight line.

\end{itemize}

The analysis described below was done with the set of filaments perpendicular to the line 
of sight with the additional following criteria:

\begin{itemize}
    \item {\bf T14 filaments:} tolerance angle $=20^{\circ}$, elongation $= 0.7$, and $RMS=0.06$.
    \item {\bf P19 filaments:} tolerance angle $=30^{\circ}$, elongation $= 0.7$, and $RMS=0.14$.
    \item {\bf M16 filaments:} tolerance angle $=30^{\circ}$.
\end{itemize}

There is a strong correlation between the parameters just shown (i.e. {\it RMS}
and elongation), we explore the use of both, to clean the samples from irregular shaped filaments. 

The most restrictive is elongation and accounting the results of
smaller {\it RMS}, therefore we only use the elongation criteria to limit our samples.

The distribution of elongations of T14 filaments is quite different than that of P19
filaments. The former has an elongation distribution with a sharp peak very near to $1$, and the 
minimum limit of $0.7$ almost does not change the filament set.
On the other hand, P19 distribution of elongations is wide, ranging from $0.5$ to $1$, and the 
minimum limit of $0.7$ considers 
the $\approx 86 \%$ of filaments. We filtered out all filaments with angular size above $4$ degrees 
to avoid contamination from close filaments.
As we use a galaxy sample limited to a maximum co-moving distance of $350~ {\rm Mpc}$, to avoid 
doing statistics with filaments above this limit, we filtered them by
leaving the sample of filaments closer to this value.

\subsection{Filament length}

We study global properties of filaments choosing 
the galaxies inside a region around its axis.  This region could be thought as
the combination of $N - 2$ balls of radius $R$ centred at the inner path nodes,
connected by $N - 1$ cylinders of the same radius and subtracted the balls of
the same radius at the end and start of the filament likely related to groups
or galaxy clusters.  
We define the filament radius $R$ as half its length for the ones shorter than $15~{\rm Mpc}$.
For those longer than this limit, we fix $R$ to $7.5~{\rm Mpc}$.  
The volume of each filament was estimated with $V = \pi R^{2} l$ where $l$ is the filament 
length, although this is not the exact formula for the filament volume, it is a good estimation 
since we selected straight filaments to study.  All galaxies in the region define the properties of 
the filament, for example the total luminosity. It is worth noticing that according to this 
membership definition, it is possible that a galaxy could be assigned to more than one filament.  
The algorithms were applied to both, real and random galaxies, to account how
different a filamentary region is to a random distribution of galaxies,
according to each catalogue. This helps us to understand what galaxy overdensities
the different algorithms  are able to find.

\begin{figure}
\centering
\includegraphics[width=0.5\textwidth]{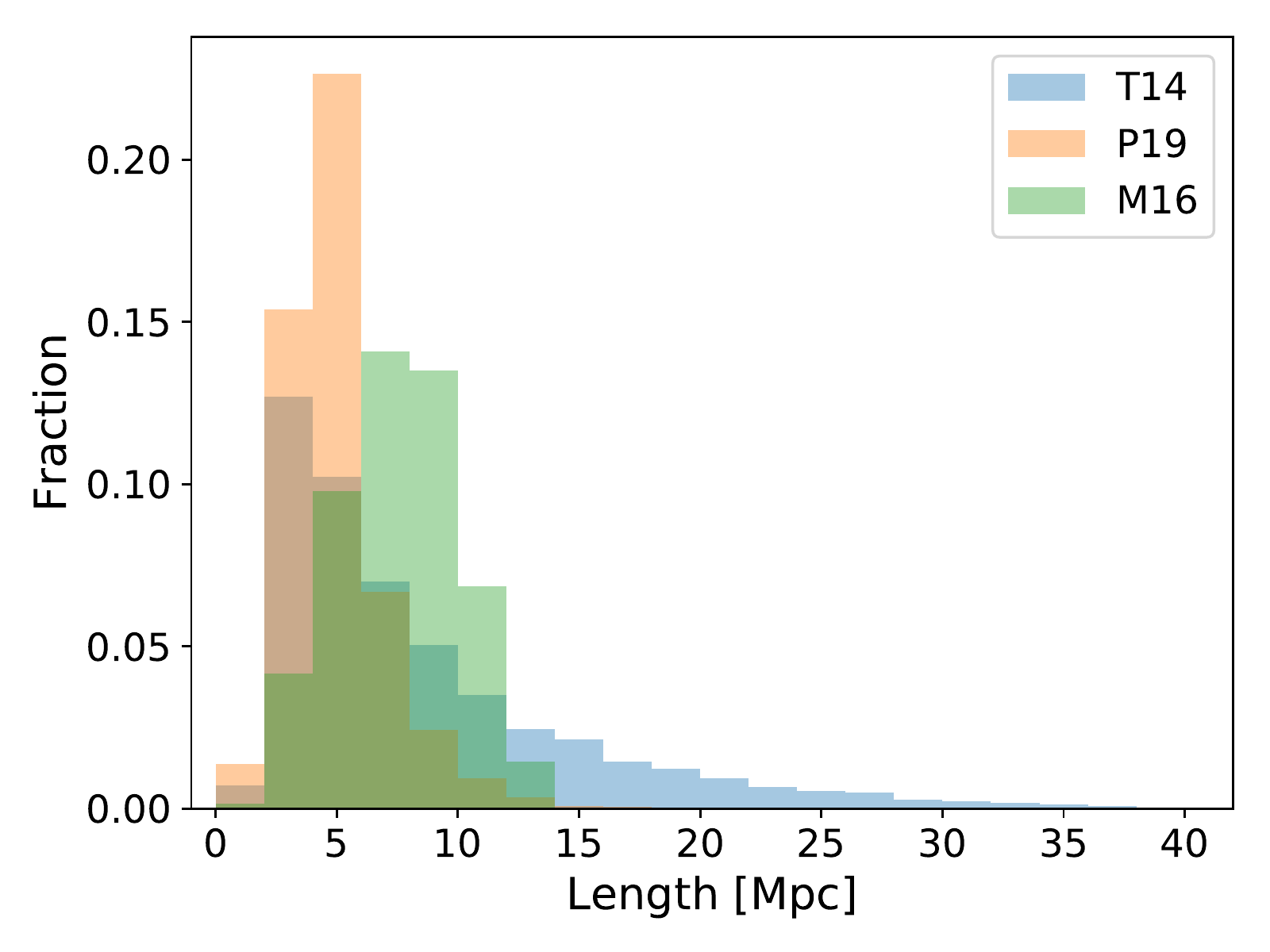}
\caption[Histogram of length.] 
{Distribution of filament length. 
The shortest correspond to the filaments of 
M16, and the largest to the T14 catalogue.}
\label{hist_length}
\end{figure}

As shown in Fig. \ref{hist_length}, the bulk
of filaments in all catalogues is located between the $5 - 10~{\rm Mpc}$ range.
However, M16 filaments are limited by $12~{\rm Mpc}$, and P19 filaments
extend this limit to $15~{\rm Mpc}$ with few exceptions larger than $25~{\rm Mpc}$. These
different ranges are explained by the fact that different algorithms (as well
as different definitions of filaments) are being used. 

\subsection{Galaxy overdensity in filaments}
In this subsection we analyse the galaxy overdensity in filaments which we compute
as:

\begin{eqnarray}
S_{\rm real} &=& \frac{N_{\rm real}}  {N_{\rm rand1}}\\
S_{\rm rand} &=& \frac{N_{\rm rand2}} {N_{\rm rand1}}
\end{eqnarray}
where $N_{\rm real}$ is the number of real galaxies
around filaments, $N_{\rm rand1}$ and $N_{\rm rand2}$ are the number of galaxies around these objects
but instead, from $2$ independent random galaxy catalogues.

The overdensity was calculated to both a real and random samples 
to estimate their distribution functions and how overlapped they were. 
The regions where the algorithms detect filaments are over-dense, therefore
the values for this parameter are above $1$ for all the catalogues.
But they may differ naturally as a consequence of the different algorithms, 
for example the P19 algorithm uses a FoF algorithm 
to discard the low density regions, before actually detecting the
filaments, while M16's defines a filament that is formally the 
line that joins two groups through an over-dense region.  
These then are mechanisms that indirectly increase
the overdensity of the filaments.

\begin{figure}
\centering
\includegraphics[width=0.5\textwidth]{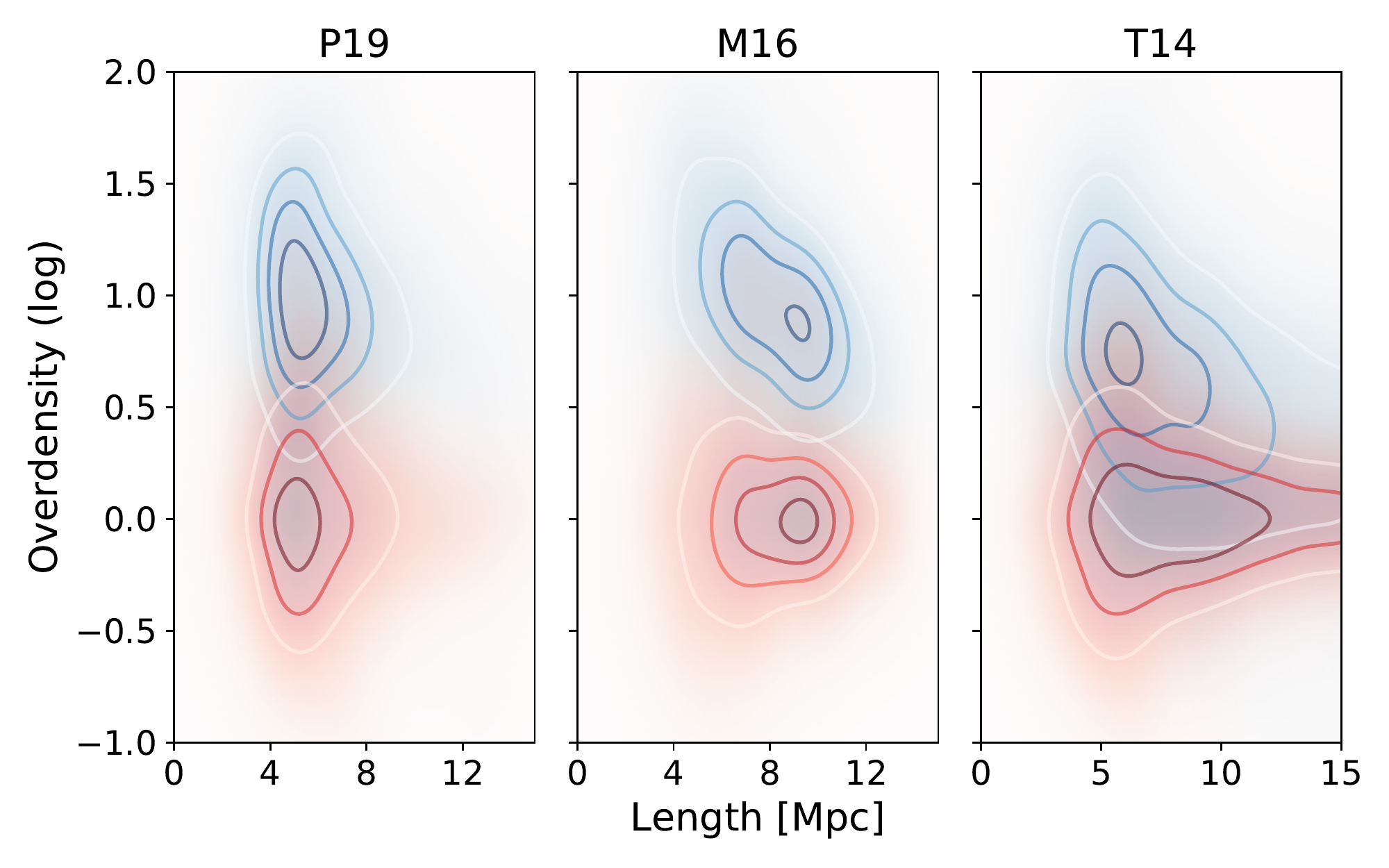}
\caption[Length vs overdensity]
{Relation between length and overdensity of filaments for the three
catalogues, in white to blue real galaxies and in white to red random galaxies. 
On the left:
filaments from P19 catalogue.  Centre: filaments from M16's
catalogue. Right: filaments from the T14 catalogue.}
\label{largo_sobredensidad} 
\end{figure}

As shown in Fig. \ref{largo_sobredensidad}, the overdensity is
approximately $10$ for the filaments in the catalogues of P19 and M16, while
for the T14 catalogue it is near $5$, compared to their respective random overdensities. 
This means that these regions contain from $5$ to $10$ times the amount of galaxies compared 
to the amount they would have with random galaxies. 

Furthermore, the overdensity tends to decrease for long filaments 
for the catalogues T14 and P19, the correlation is difficult 
to see in the case of the catalogue M16 due to the limited length range.
This could mean that long filaments do actually have less density 
because short filaments are nearer to galaxy clusters, 
or, alternatively, that the radius of those filaments is so large that considers
regions close to the filament that are not related to the filament itself, thus
lowering the overdensity. 
If this is the case, another radius as a function of the length has to be considered.

\subsection{Filament luminosity}

\begin{figure}
\centering
\includegraphics[width=0.5\textwidth]{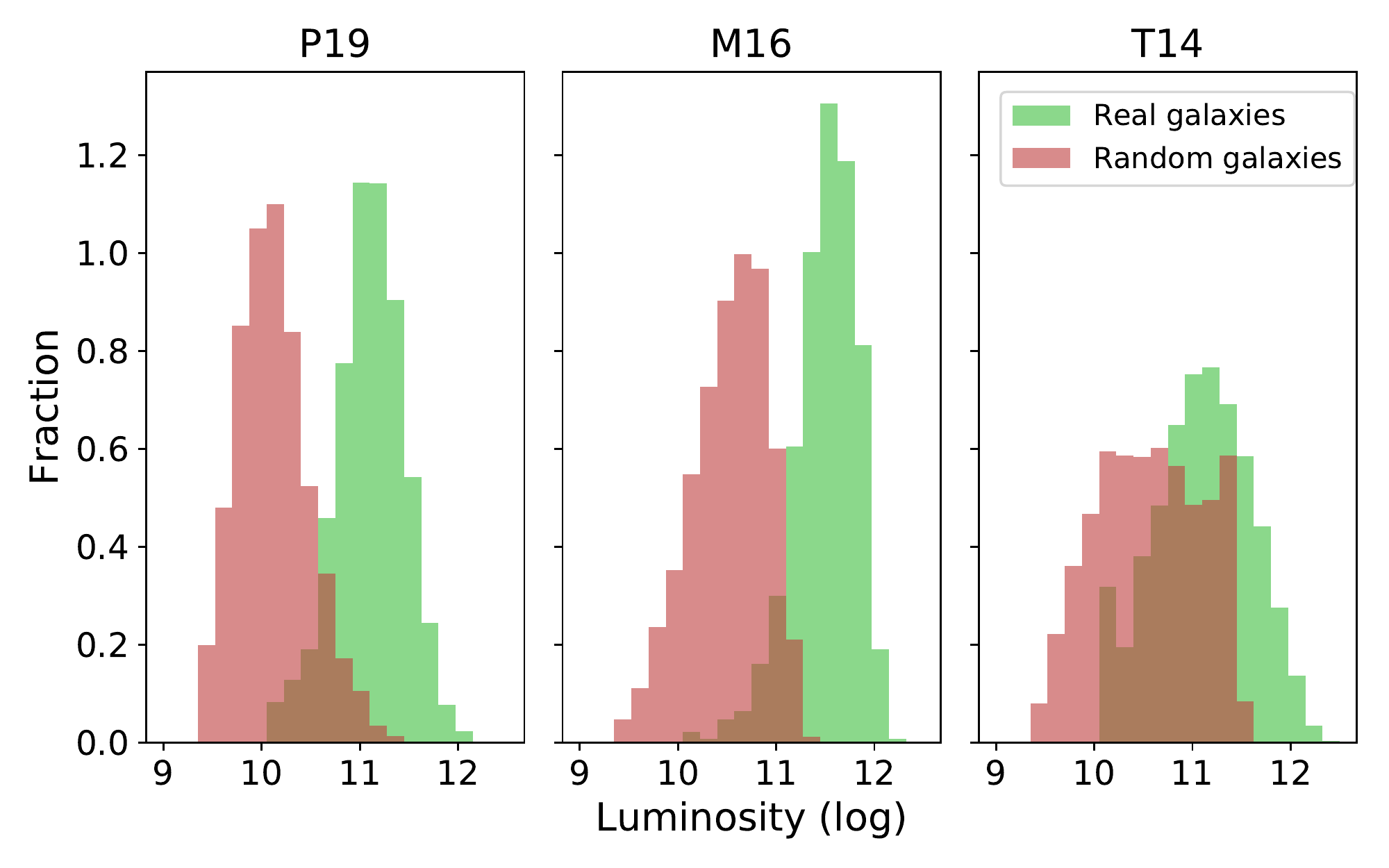}
\caption[Histogram of length.]
{The distributions of galaxy luminosity for real and random galaxies, in colours green and red, 
respectively. 
On the left: filaments of the P19 catalogue.
In the centre: filaments of the M16 catalogue. 
Right: filaments of the T14 catalogue.}
\label{hist_lum}
\end{figure}

The histograms of luminosity of each catalogue can be seen in Fig. \ref{hist_lum}.
It is worth noticing that the three catalogues find filaments in regions with higher 
luminosity than average. The greatest difference between the random and real galaxies, 
are seen in M16 and P19 samples, with similar shapes in the distribution 
and a clear difference in the average luminosity between them. For the T14 sample, 
both distributions overlap, random galaxies have a broader distribution and real galaxies have a sharper 
peak.  The luminosities are about $10^{11.41}L_{\odot}$ for real galaxies, and
$10^{10.93}L_{\odot}$ for random galaxies, for T14 filaments.
For the P19 sample the quantities are $10^{11.22}L_{\odot}$ for real galaxies, and
$10^{10.31}L_{\odot}$ for random galaxies. Finally, for the M16 sample, the values are
$10^{11.55}L_{\odot}$ for real galaxies, and $10^{10.56}L_{\odot}$ for random galaxies.
A similar tendency is found in the galaxy number's histogram (not shown),
however there is a higher difference in the distribution for M16 and
P19 and quite similar for T14.  The average values of the samples
are $11.9$ galaxies in T14 filaments, $7.23$ for P19 and $15.1$ for
M16.

\subsection{Mean galaxy luminosity in filaments}

\begin{figure}
\centering
\includegraphics[width=0.5\textwidth]{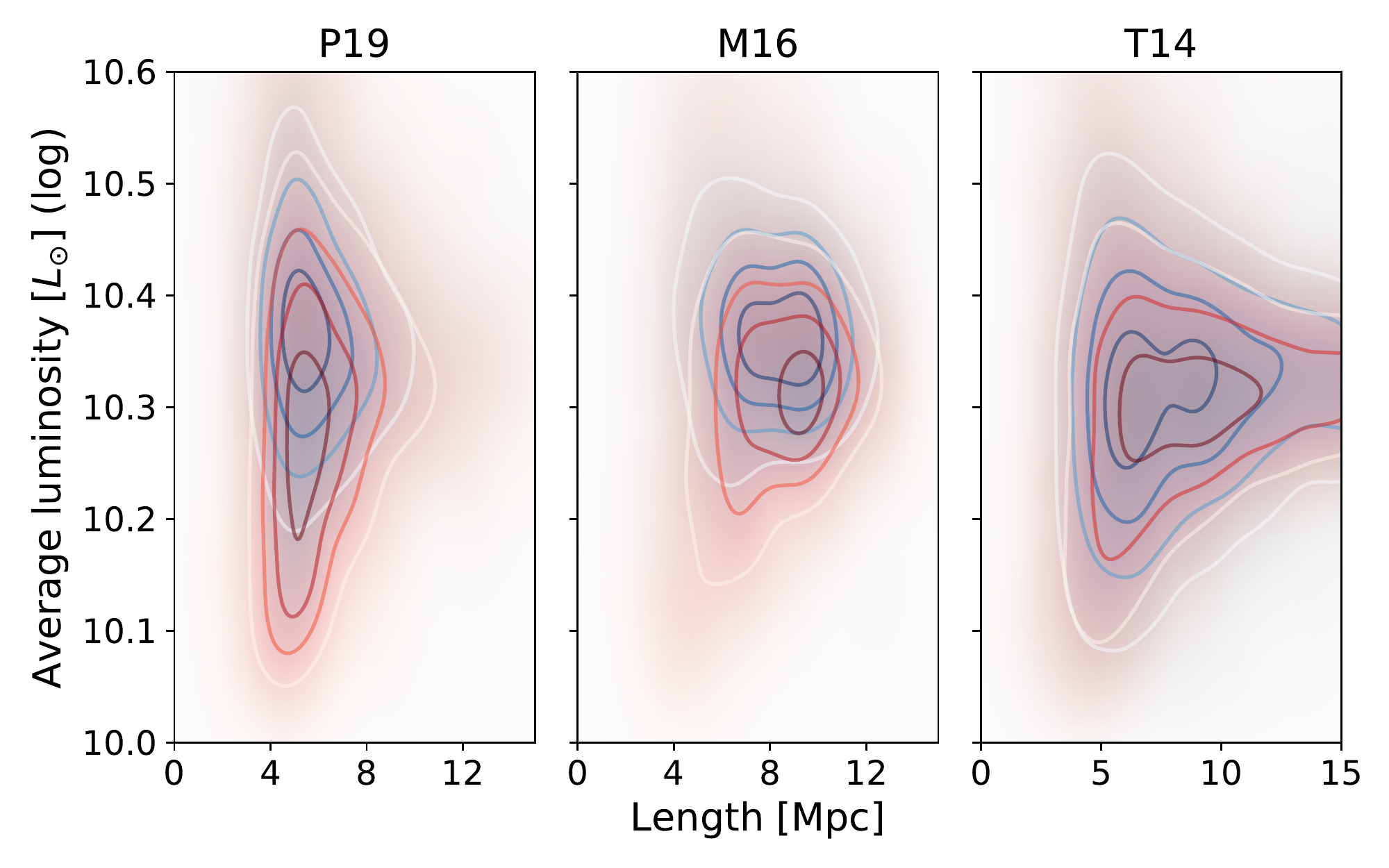}
\caption[Length vs overdensity]{
Relation between length and average galaxy luminosity of filaments for the
three catalogues, in white to blue real galaxies and in white to red random galaxies.  
On the left: filaments from the P19 catalogue.  Centre:
filaments from M16's catalogue. Right: filaments from the T14 sample.
}
\label{largo_brillo} 
\end{figure}

Another parameter we focus on is the average galaxy luminosity in filaments as
the mean of the individual galaxy luminosities in the filament 
(see Fig.\ref{largo_brillo}). It is not possible to find a correlation
between the length and the average luminosity in general, not even a strong
difference between the random and real samples.  However, the difference is
less significant in Temple's filaments, being a little more luminous the real
galaxies (compared with the random samples) for the other two catalogues.  For
P19 case this is not surprising, because there 
always exists a path of luminous galaxies between the extremes.  
In general we see a larger dispersion
in the luminosities for short filaments, this is most likely by the effect of
low number statistics, as the effect can be reproduced with the
random galaxies.

\subsection{The luminosity function of galaxies in filaments}
We compute the luminosity function (hereafter LF) of galaxies in subsamples of
filaments defined by their length. Since in this work we use volume limited samples of 
galaxies  our computations of the LFs are restricted to the absolute magnitude range  
$-23.0 \le M_r \le -20.5$, i.e., our LFs are probing only the bright end of the
LF. Therefore, when we compute best fit parameters of the \citet{Schechter:1976} 
function below, the $\alpha$ parameter is related to the shape of the bright end of 
the LF, and does not measure the faint end slope of the LF as is the usual case in the
literature. We restrict the LF computations to galaxies in our tracer sample 
in the redshift range common to the three filament catalogues: $0.05\le z\le0.137$.

\begin{table*}
    \centering
    \begin{tabular}{lccccccccc}
                 & \multicolumn{3}{c}{All sizes} & \multicolumn{3}{c}{$2-6$ Mpc} & 
                 \multicolumn{3}{c}{$6-10$ Mpc} \\      
                 \hline
                 & $N_{\rm gal}$ & $\alpha$ & $M^{\ast}$ & 
                   $N_{\rm gal}$ & $\alpha$ & $M^{\ast}$ & 
                   $N_{\rm gal}$ & $\alpha$ & $M^{\ast}$ \\
                 \hline
      Field     & 102,832 & $-0.62\pm0.06$ & $-20.48\pm0.03$ & 
                      $-$ & $-$ & $-$ & 
                      $-$ & $-$ & $-$ \\
      Groups    &  45,701 & $-1.04\pm0.04$ & $-21.19\pm0.03$ & 
                      $-$ & $-$ & $-$ & 
                      $-$ & $-$ & $-$\\
      P19       &  6433   & $-0.26\pm0.02$ & $-20.58\pm0.01$ & 
                   3287   & $-0.26\pm0.02$ & $-20.64\pm0.02$ & 
                   2314   & $-0.26\pm0.08$ & $-20.54\pm0.06$ \\
      M16       &  5259   & $-0.6\pm0.1$   & $-20.6\pm0.1$   & 
                    654   & $-0.7\pm0.4$   & $-20.8\pm0.3$   &
                   4429   & $-0.5\pm0.1$   & $-20.6\pm0.1$   \\
      T14       & 47,946  & $-0.77\pm0.07$ & $-20.66\pm0.04$ & 
                    5740  & $-0.9\pm0.2$   & $-20.8\pm0.1$   &
                  11,052  & $-0.8\pm0.2$   & $-20.70\pm0.08$ \\
                 \hline
    \end{tabular}
    \caption{Luminosity function of galaxies in the field, in groups and in the three samples of filaments: best fit Schechter's function parameters to the LFs in the absolute magnitude range $-23.0 \le M_r \le -20.5$, obtained using the STY estimator.}
    \label{tab:lf}
\end{table*}

For galaxies in filaments we consider the overall LF of the three samples, and also
subsamples of filaments of different redshift space length: $2-6$, and $6-10$ Mpc.
We also compute the LF in groups, and in the field, to compare with the samples of
filaments. We construct a sample of galaxies in groups by identifying galaxy groups and 
clusters using a modified FoF algorithm as described in \citet{Merzan2005} with a transverse
linking length corresponding to an overdensity of $\delta\rho / \rho =200$ 
and a line-of-sight linking length of $V_{0} = 200\ \mathrm{km\ s^{-1}}$. 
We restrict our analysis to massive, $\log(M_{\rm vir}/M_{\odot})\ge 13.5$,
groups, i.e., those used in the construction of the M16 filament sample.
Our sample of field galaxies comprises all galaxies in the volume under study 
that are not assigned to groups by the FoF algorithm, nor to filaments by any of 
the filament samples we use.

\begin{figure}
\centering
\includegraphics[width=0.5\textwidth]{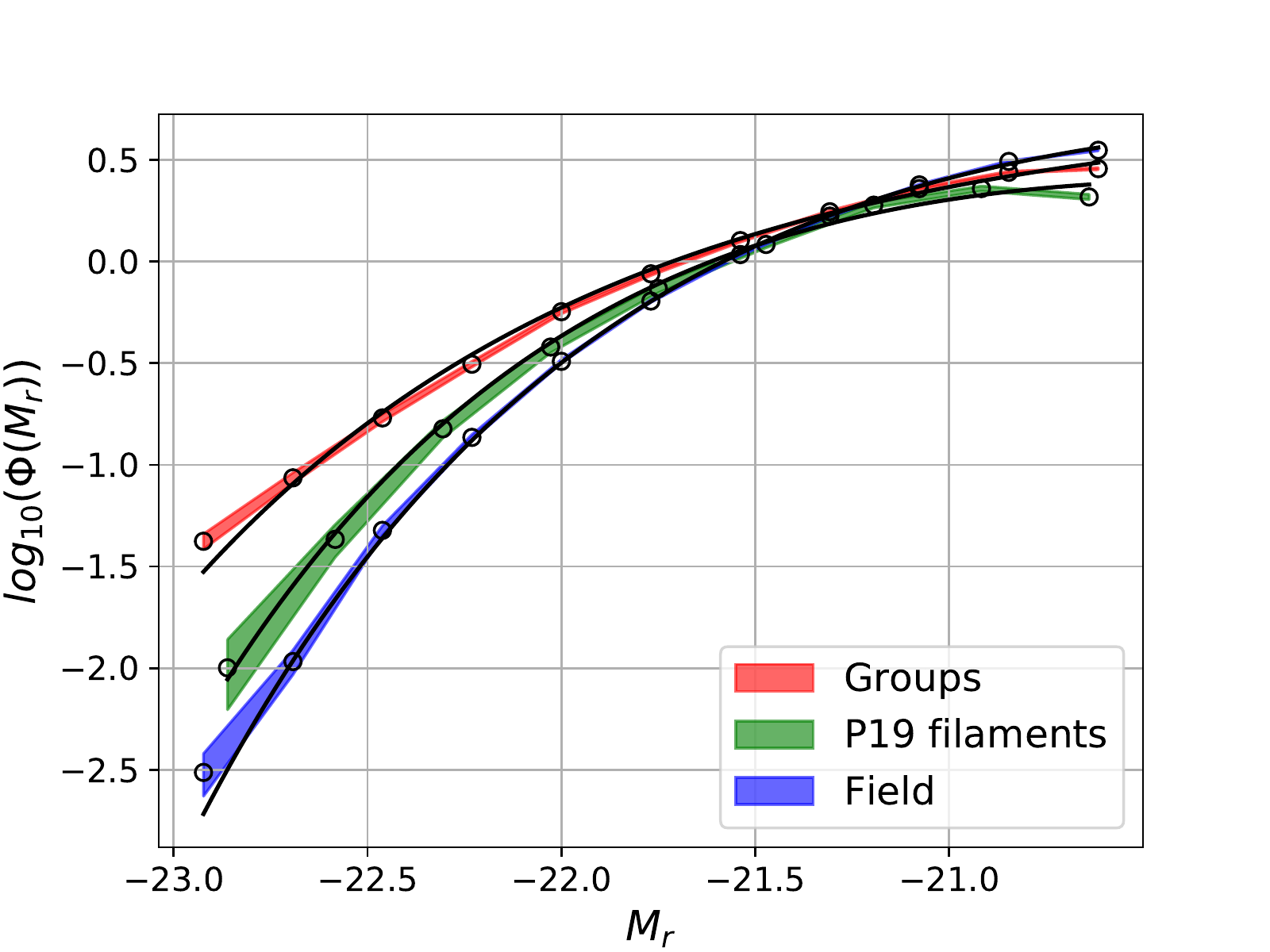}
\caption[Luminosity functions.]
{Three examples of the galaxy luminosity functions in different environments. Points and error-bars are computed using the SWML method. We show the best-fit Schechter functions with 
parameters computed using the STY estimator. 
Curves were scaled up/down for comparative purposes.}
\label{lum_functions}
\end{figure}

We use two standard methods to compute the LF: 
the step-wise maximum likelihood (SWML, \citealt{SWML}) to produce binned LF, 
and the parametric STY estimator \citep{STY} to compute the best-fit Schechter function 
parameters. We do not attempt to compute the normalisation of the LF,
our interest is to study differences in the characteristic magnitude and the shape of the
LF between the different subsamples of galaxies.
We show in Fig. \ref{lum_functions} examples of the LFs we compute, 
along with their corresponding best fit Schechter functions. In all cases we study, the
best fit Schechter model is a good description of the binned LF, therefore,
we focus our discussion in the comparison of the parameters $M^{\ast}$, and $\alpha$,
across samples.

We quote in Table \ref{tab:lf} the resulting best fit Schechter parameters of the different
LFs we compute, as obtained using the STY estimator. 
On the one hand, the LF of galaxies in groups has the brightest $M^{\ast}$, and the
smallest $\alpha$.  On the other hand, the LF of field galaxies has the dimmest
$M^{\ast}$ of all the samples.
In a qualitative agreement with the results of \citet{martinez}, we find that
the characteristic magnitude $M^{\star}$ of the LF of galaxies in filaments takes, 
in all cases, an intermediate value between those of field and galaxies in groups, 
but closer to field values. The $\alpha$ parameters of the samples M16 and T14
are also intermediate between the corresponding parameters of field, and group galaxies.
However this is not the case of P19, whose $\alpha$ parameter is the largest in all 
cases.

In the comparison between the LFs of galaxies in filaments in the three catalogues,
there are a few points to consider:
\begin{enumerate}
\item The $\alpha$ parameter of P19 LFs is much larger than those of the other
two filament samples in all cases, and seems to be independent of filament size.
\item Both parameters of the LFs of M16 and T14 are, for all filament sizes, consistent
within error-bars. 
\item The $M^{\ast}$ values of the T14 LFs are systematically brighter than those of P19.
A similar conclusion can not be drawn when comparing the $M^{\ast}$ values of M16
and P19 given the large error-bars in the M16 parameters, due to the smaller sample.
However the tendency is for the M16 sample to have brighter values of $M^{\ast}$.
\item A tendency of shorter filaments to have brighter $M^{\ast}$ is seen, this is, however,
statistically significant only for the P19 sample.
\end{enumerate}

Despite the differences in the construction of the samples by T14 and M16, their
LFs are similar. P19 filaments, on the other hand, have a distinct LF characterised 
by a much higher value of $\alpha$. This consistency of P19 filaments to have, in all 
cases, the largest values of $\alpha$, makes them more unlikely to host galaxies in the 
bright end than the other two samples.
Recall that we are probing the bright end of the
LF and $\alpha$ is a measure of the convexity of the LF in this magnitude range, and
not a measure of the faint end slope, which our samples of galaxies do not probe.

All the characteristics shown in this section suggest that properties of 
each catalogue strongly depend in the way they were build.
\citet{Tempel2014} algorithm is based on geometrical and stochastic assumptions of the
distribution of galaxies, whereas \citet{martinez} algorithm detects overdensities
between galaxy groups and \citet{pereyra_submitted} method finds filaments as paths of
luminous galaxies.

\section{Spatial galaxy distribution} \label{sec:analysis}

In this section we continue our analysis on the differences between the samples
of filaments T14, M16, and P19, by focusing on a number of spatial features of the
samples. We use a spatial stacking scheme that we detail below, and compute an
adaptation of the two point correlation function that measures the  projected clustering
of galaxies along the direction perpendicular to the filament axis.

\subsection{Spatial Filament Stacking}
\label{sec:stacking}

We stack data of several filaments to enhance the information that
different populations of these objects have, increasing 
the signal/noise ratio.  We proceed as follows:
Firstly, we define a base set of filaments from each filament catalogue, in which bend
and too short filaments are filtered out (see section \ref{sec:GralProp}),
resulting in a set of straight filaments. 
Then, for each filament and its surroundings, we define two plane-of-the-sky projected 
Cartesian coordinates, one alongside the filament direction and centred
in one of its extremes ($x$), and the other perpendicular
to it ($y$). Since every filament has a different length, we re-scale both
coordinates in order to have filament length equal to 1. 
This results in having all filaments with their starting and ending points 
at the coordinates $(0,0)$ and $(1,0)$ respectively.
With this normalisation each part of the
filaments, such as the start and end (associated normally with galaxy groups or
clusters), the middle filamentary part, the signal beyond the nodes
(associated with correlation between connected filaments) and the rest of the
field will be stacked at the same places. If no normalisation is done,
short/long and far/close filaments will be stacked with different sizes and the
galaxies from different parts of each filament will mix together.  This
procedure is repeated using galaxies from the random catalogue of galaxies.

To process each field near the filament, the angular length of the filament is
measured to select an area large enough to cover all the filament field that
will be stacked. To avoid summing all the data along the visual axis, that is
uncorrelated and adds noise, galaxies with distances further than $10~ {\rm Mpc}$
 from both the start and end of the filament are not
considered.  The criteria used to consider filaments are the same of the
section \ref{sec:GralProp}.

\begin{figure}
\centering
\includegraphics[width=0.90\columnwidth]{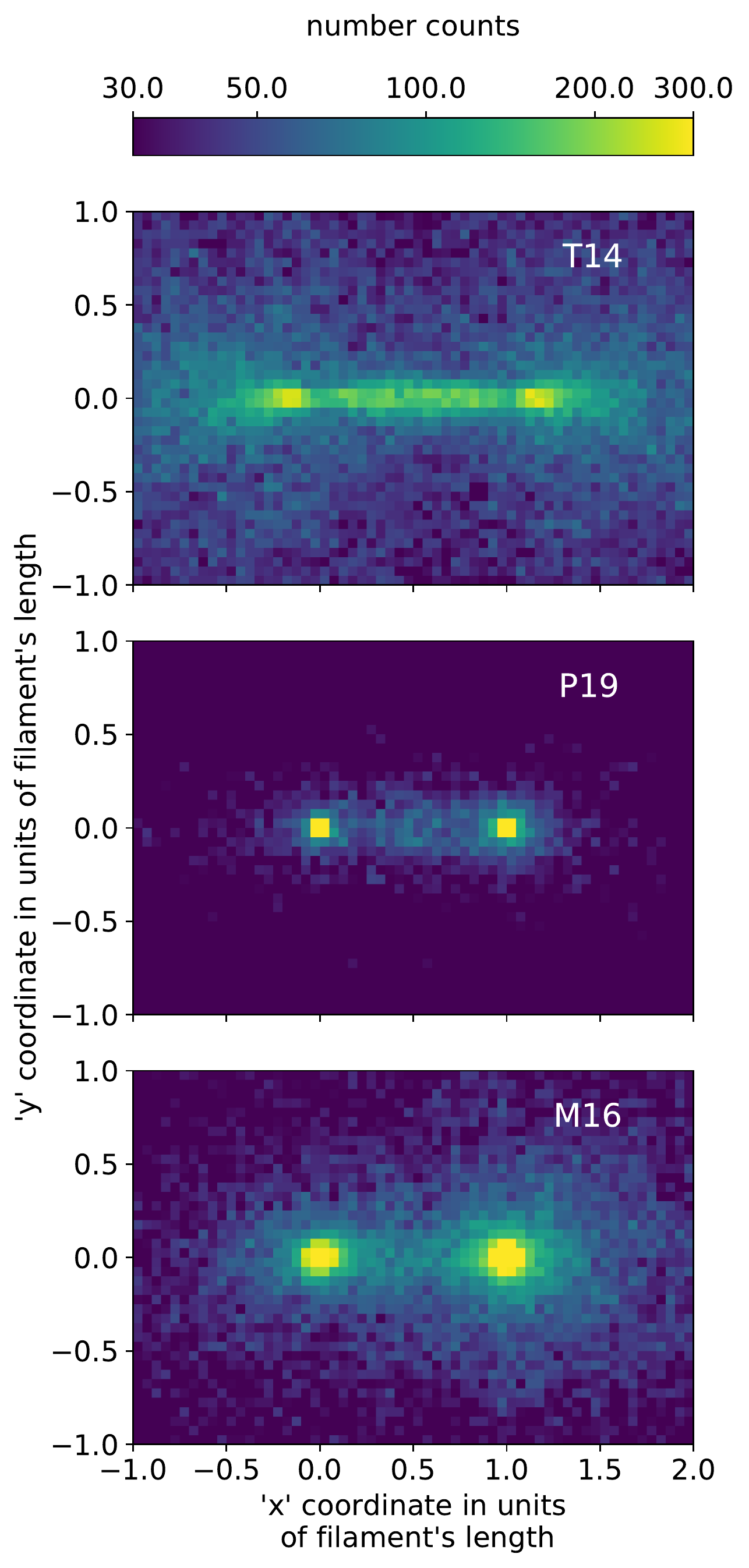}
\caption[2d Histogram.]
{
Two-dimensional histogram of the filament stacking. Colour indicates the galaxy count. 
The filament shape show different features for all catalogues,
M16 and P19 filaments have clear node radial distributions and T14
have a strong signal and the peaks at nodes are shifted.} 
\label{hist2d} 
\end{figure}

There is evidence of filaments that are not necessarily connecting
two high density peaks (the so-called {\it tendrils } in \citealt{Alpaslan2014}).
However the only algorithm capable of identifying similar structures is the one used to build P19, and for this catalogue, a selection criteria was used to filter them out.
In Fig. \ref{hist2d} we show the resulting stacking procedure.
Each sample of filaments has a different profile but it is possible to distinguish the
typical shape of a filament with $2$ peaks of density at the extremes,
indicating the average position of the galaxy groups or clusters, and a high
density filamentary region connecting those extremes. 
In the case of the M16 catalogue, the extremes have a perfectly radial 
profile, this happens because
those filaments were detected as pairs of galaxy groups. On the other hand,
P19 filaments have high density peaks at the extremes because, by construction, there are
always bright galaxies at the extremes, and at least one galaxy in the path
between them. With this definition, it is natural that there is a path of
high density matching the extremes. It is noticeable also the presence of signal 
presumably from adjacent filaments beyond the nodes for the case of T14.
In general it is observed a high symmetry
in both $x$ and $y$ axis with M16 and T14 filaments, in the case of
P19, the most luminous extreme is always placed on the right, so there is
less symmetry in the $x$ axis. 

\subsection{Mean galaxy overdensity profile of filaments}
We build a one-dimensional profile that measures the overdensity of galaxies as a function of the
distance to the filaments' axis.
We consider the region defined by $0.15 \leq x \leq 0.85$, $-1 \leq y \leq 1$,
and define the mean overdensity profile as:
\begin{equation} 
I(|y|) = \frac{R(|y|)}{A(|y|)} - 1, 
\label{I_de_y}
\end{equation}
where $R(|y|)$ is the number of real galaxies, and $A(|y|)$ the normalised number of random galaxies,
at a distance $|y|$ from the filament axis. 
The error-bars of these functions are calculated with the Jackknife method, dividing the
sample into $N/2$ subsamples (where $N$ is the number of elements of the sample), 
therefore calculating each computation excluding two of the filaments and determining  
the uncertainties.
We expect the overdensity profile to reach a maximum near $y = 0$,
while at large values of $|y|$ the signal should vanish and $I(|y|) \approx 0$.

\begin{figure}
\centering
\includegraphics[width=0.5\textwidth]{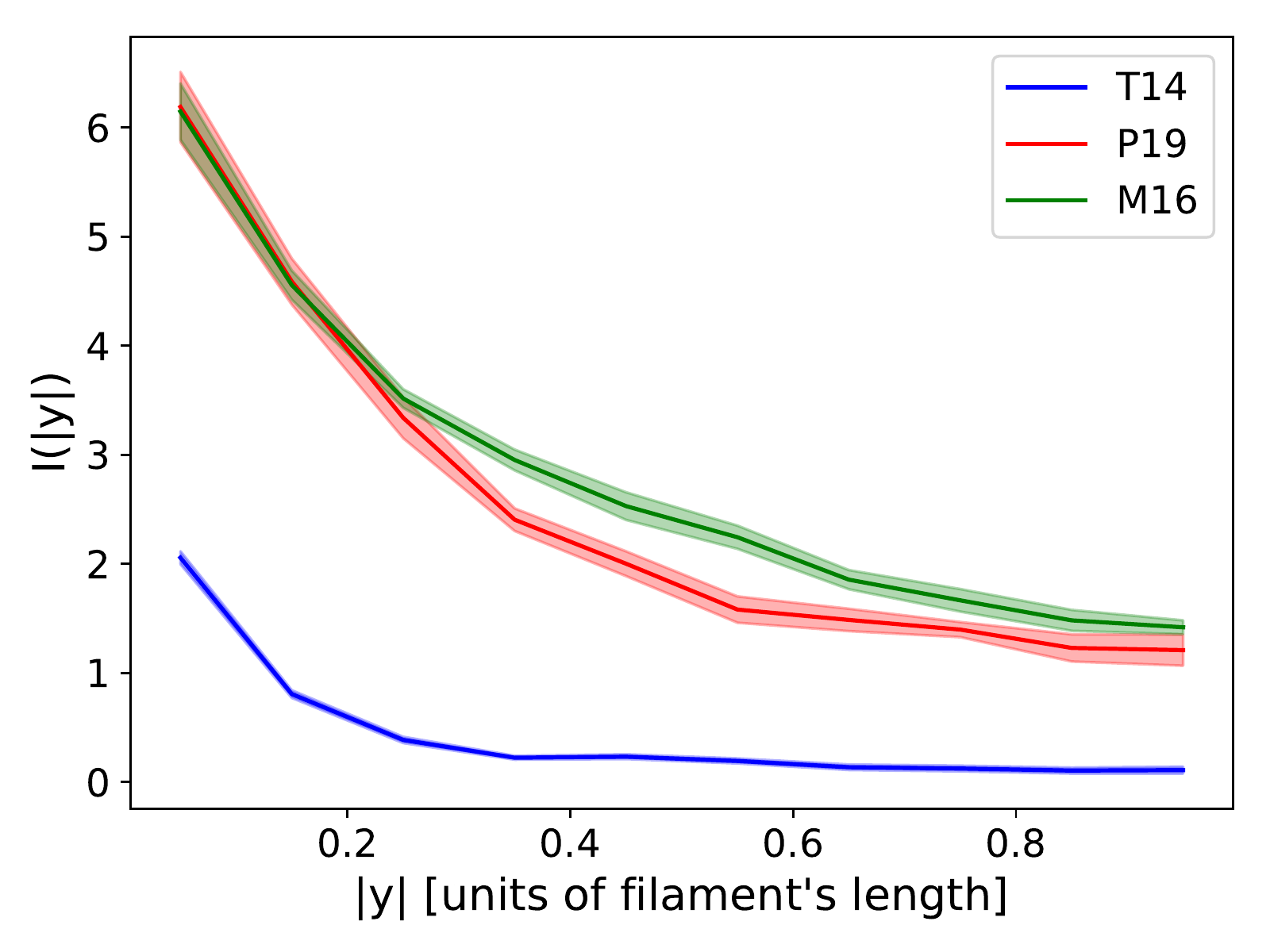}
\caption[2d Histogram.]
{Mean overdensity profile of galaxies as a function of the normalised distance to the
filaments' axis, as defined in Eq. \ref{I_de_y}.
Colours represent different samples of filaments as shown in the inset plot.
}
\label{stacking_total}
\end{figure}

In Fig. \ref{stacking_total} we show the overdensity profiles for the three samples 
of filaments, which are consistent with the stackings shown in Fig. \ref{hist2d}.
As expected, filaments are over-dense regions defined by the large scale structures.
In general, the maximum overdensity is in the same magnitude order for all catalogues.  
The signal is stronger for M16 and P19 because they were constructed considering physical 
objects as points that define the filament path.
On the other hand, the T14 catalogue reaches a much lower value and its profile
decreases steeper to the background than the other catalogues.

\begin{figure}
\centering
\includegraphics[width=0.5\textwidth]{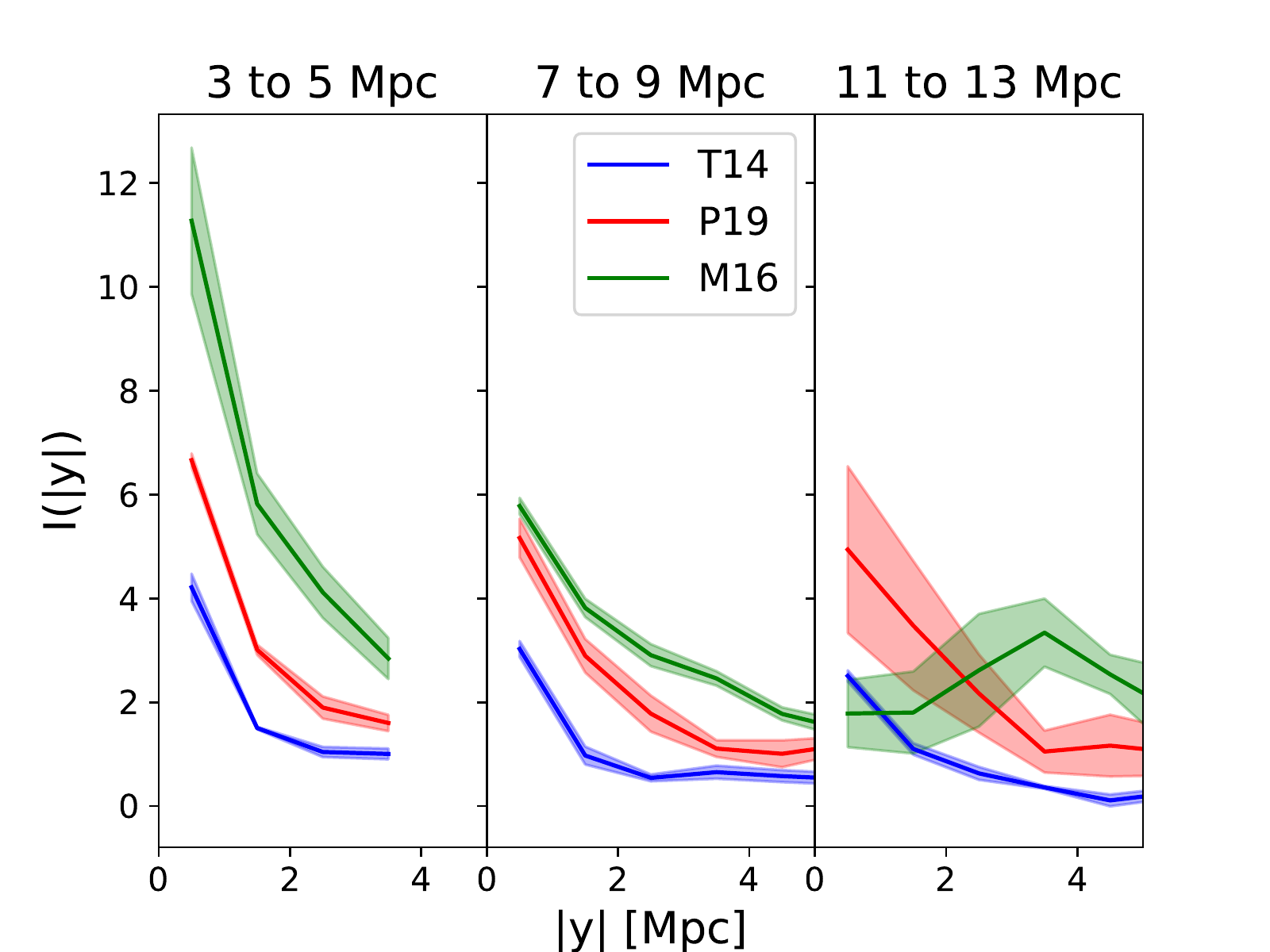}
\caption[2d Histogram.]
{Mean galaxy overdensity profile around filaments split in three bins of
filament length, as indicated on top of each panel.
Note that short filaments have a much higher overdensity profile
than long ones.}
\label{stacking}
\end{figure}

We study how the overdensity profile is related to filament length by dividing the samples 
into three different sets per catalogue separating the filaments by length. 
The bins in filament length we consider are: $4\pm1$, $8\pm1$, and $12\pm1 ~{\rm Mpc}$. 
In Fig. \ref{stacking} we show these profiles and it is possible 
to observe a general decrease of the mean overdensity when increasing the
filament's length, it is also observed in Fig. \ref{largo_sobredensidad}, this is in agreement with 
the idea of that {\it strong} 
filaments tend to be short bridges that match close galaxy clusters, while for 
further clusters, filaments are weaker in general \citep{bond}.
We find that there is a strong variation for M16.
The other catalogues show a milder trend. 
This fact can be justified again in the way the filaments were constructed. 
As M16 catalogue is constructed from groups, 
is understandable that they have a larger correlation for shorter filaments.

\subsection{The overdensity profile of blue and red galaxies}

It is possible to use different types of galaxies as tracers of the overdensity profile 
to study the properties of the filaments.
If we separate by colour, it is expected that the red galaxies have higher overdensities at the 
centre of the filaments, in contrast with
the blue galaxies that tend to locate around the filaments as has been shown by \citet{kraljic}, 
and similarly in the works of \citet{dressler} and \citet{blanton} with galaxy clusters.
We also expect that the distributions of red and blue galaxies are different whether they are close to galaxy clusters or groups (short filaments) or far away from them (long filaments). Having the former 
a higher overdensity of red galaxies at the filament axis compared to that of blue galaxies.

We follow \citet{omill2011} and \citet{FernandezLorenzo2012}, and define red galaxies as
those that satisfy $g - r \geq 0.7$, and otherwise for the blue galaxies. 
This colour separation divides the red-blue bi-modal distribution through the
green valley. Blue galaxies comprise the $57.8 \%$ of the volume-complete
sample, this proportion is roughly constant with $z$ with a slight
tendency for lower redshift galaxies to be redder.

\begin{figure}
\centering
\includegraphics[width=0.5\textwidth]{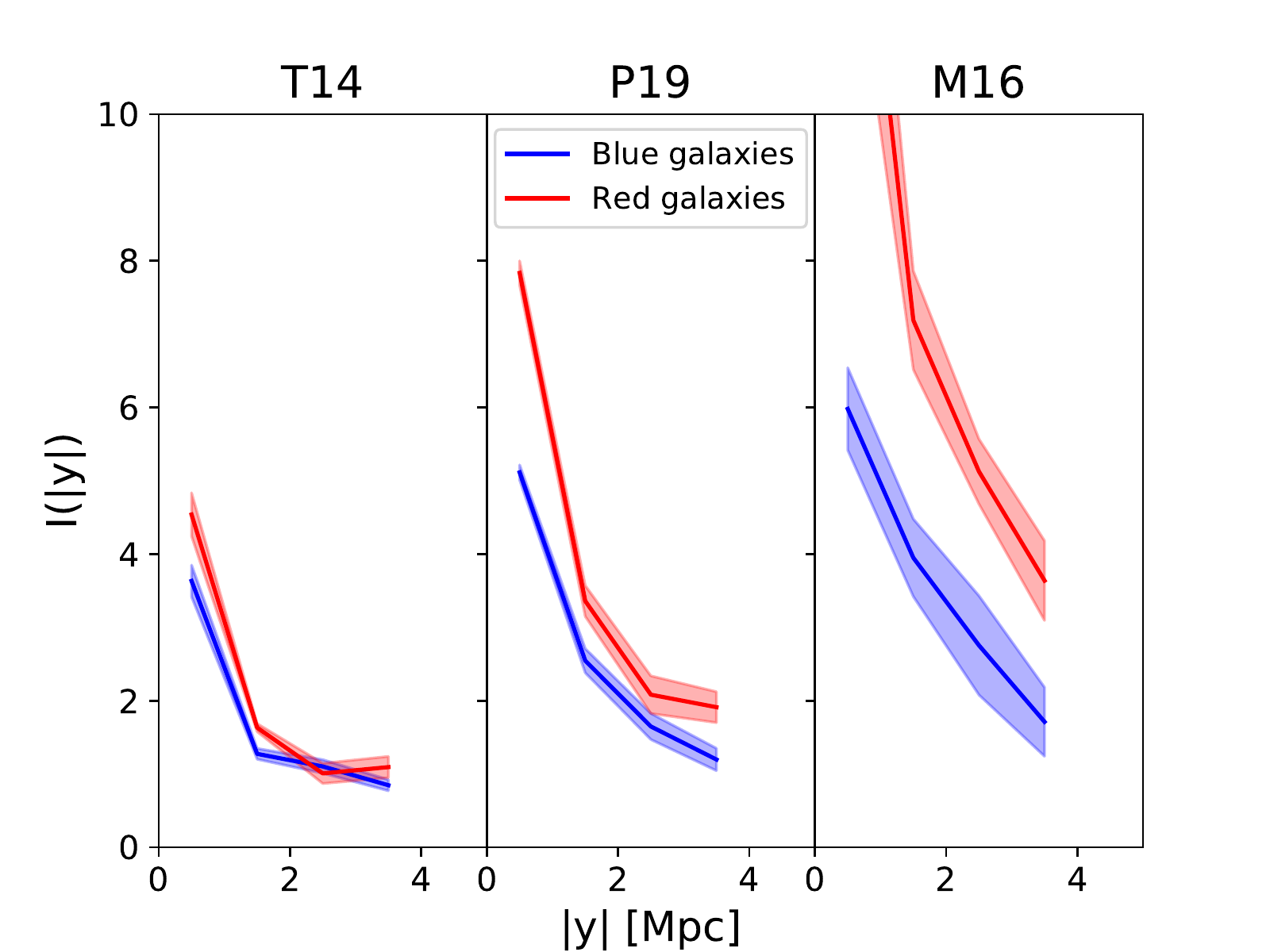}
\caption[Stacking $4 {\rm Mpc}$]
{The overdensity profile of red and blue galaxies around filaments of length 
$\sim 4~{\rm Mpc}$ from the different catalogues, as quoted on top of each panel. 
}
\label{stacking4} 
\end{figure}

\begin{figure}
\centering
\includegraphics[width=0.5\textwidth]{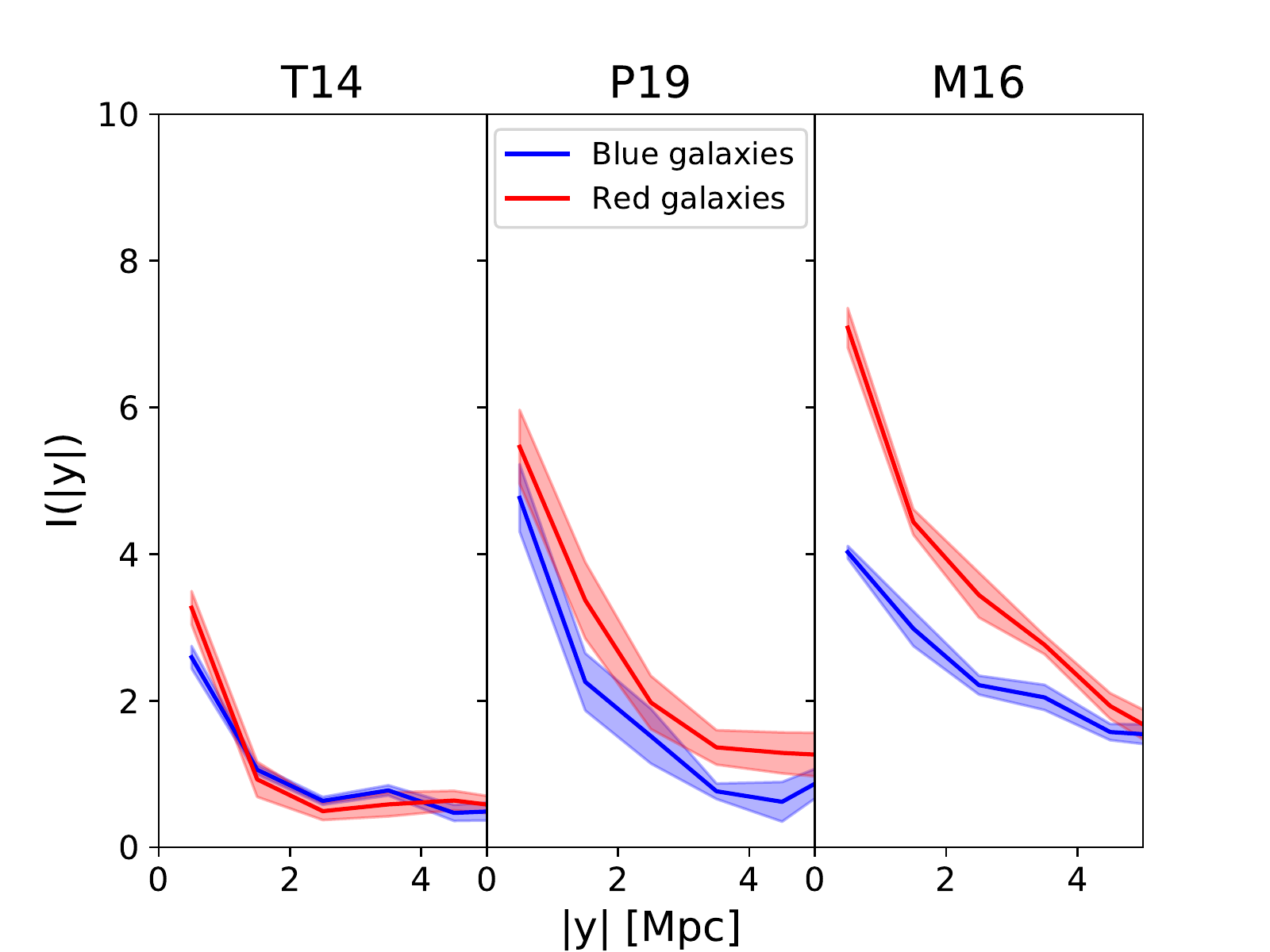}
\caption[Stacking $8 {\rm Mpc}$]
{Same as Fig. \ref{stacking4}, but for filaments of length approximately $8~ {\rm Mpc}$.
In all samples, red galaxies have a higher overdensity. }
\label{stacking8}
\end{figure}

\begin{figure}
\centering
\includegraphics[width=0.5\textwidth]{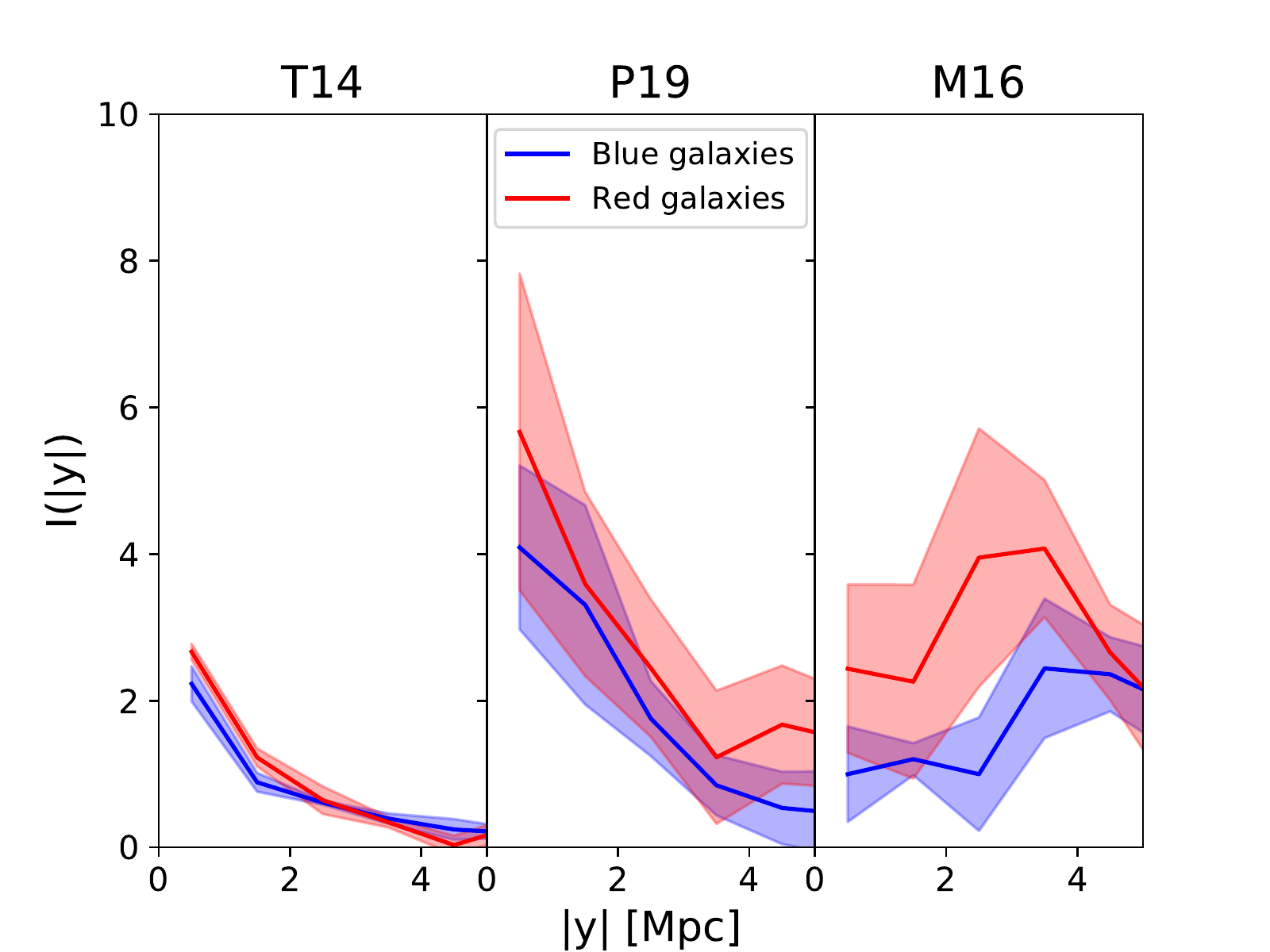}
\caption[Stacking $12 {\rm Mpc}$]
{Same as Figs. \ref{stacking4} and \ref{stacking8}, but for filaments
of length $\sim 12~ {\rm Mpc}$. Red galaxies tend to have a higher overdensity than
that of blue galaxies, however there is no statistical significant difference in any case.}
\label{stacking12} 
\end{figure}

The results are shown in Fig. \ref{stacking4} for filaments of $4~ {\rm Mpc}$,
in the three catalogues. 
In contrast with Fig. \ref{stacking}, in this figure we use physical units in distance, since the filaments considered here are similar in length.
Red galaxies show a high overdensity in the
centre of the filaments. Blue galaxies, on the other hand, still have a high
overdensity in the centre, but it is lower than the that of red galaxies.
This is in good agreement with the works of \citet{dressler} and \citet{blanton} with galaxy clusters.
P19 filaments have a higher overall overdensity, M16's follow slightly below and 
T14 have the least. This could be explained by the fact that T14 uses a stochastic
geometrical algorithm.
For filaments of $8 ~{\rm Mpc}$ (Fig. \ref{stacking8}), the overdensity decreases, and 
there is still a clear difference between the overdensity profiles of red and 
blue galaxies in the samples P19 and M16. T14 filaments do not show a strong 
difference between red and blue galaxies.
Filaments of $12 ~{\rm Mpc}$ have overdensity profiles similar to those of 
$8 ~{\rm Mpc}$ (Fig. \ref{stacking12}). 
P19 and T14 samples still maintain a slightly higher overdensity for
red galaxies. In T14 case this tendency is reverted beyond $\sim 4~{\rm Mpc}$
and M16 filaments do not show a clear signal and they are noisy, most likely because 
there are few with these lengths.

In general, all catalogues show differences between red and blue galaxies, and, 
at fixed distance, the overdensity profiles decrease with increasing
filament length, regardless of the filament sample.
T14 filaments show the least difference between red and blue galaxies, while the largest
differences are seen in the M16 case, although for this sample the profiles are noisy 
when we consider larger filaments. 
In the P19 case, the smooth overdensity profile is still present for large 
filaments ($12~ {\rm Mpc}$), however, the difference between red and 
blue galaxies vanishes. This can be understood as a consequence of the filament 
identification method itself, because P19 filaments are 
constructed through a luminous galaxy path, and therefore confusing red and blue 
in the outskirts of those galaxies.
The increase of uncertainties for the largest set of filaments may be due not only to 
the small number of filaments, but also to the internal substructure, that will tend to erase the 
difference in the relative abundance of red and blue galaxies towards the centre of the filaments.

\begin{figure}
\centering
\includegraphics[width=0.5\textwidth]{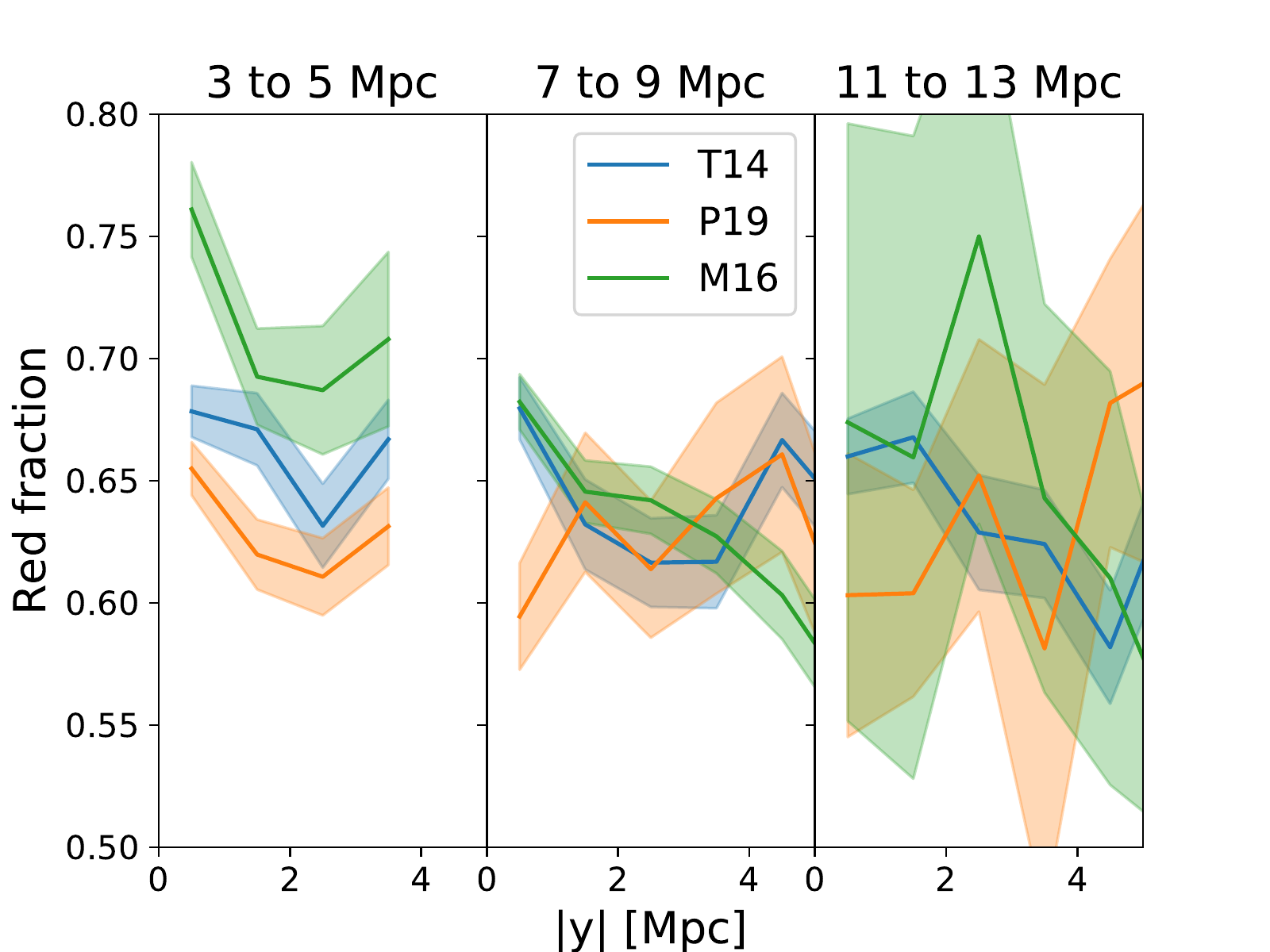}
\caption[Red fraction profiles for filaments of all catalogues.]
{Fraction of red galaxies as a function of the distance to the filament axis.}
\label{perfil_rojas}
\end{figure}

\citet{kraljic} consider filaments as {\it highways} of galaxies
that can perturb their evolution. If this were the case, galaxies near the nodes should 
have been flowing through the filament for longer time than galaxies in the centre or saddle point.
This would cause that the closer a galaxy is to the nodes (as shown by \citealt{martinez}, and 
\citealt{Salerno2019}), the redder it is.
We show in Fig. \ref{perfil_rojas} the fraction of red galaxies as a function of the distance to the
filaments' axis.  As we move from shorter to larger filaments, the fraction of red galaxies as
a function of distance is lower. Shorter filaments are expected to reside in relatively over-dense regions, 
they are expected to have preferentially red galaxies. Furthermore, it is expected that this short 
filaments are less independent of the nodes (i.e. behaving like a bridge between them), than larger 
filaments \citep{Guo2015}. For the largest filaments we analyse, this fraction
becomes noisy. Low number statistics do not allow us to study the fraction of red galaxies, as
in Fig. \ref{perfil_rojas}, but distinguishing between those that are closer to the nodes, or to the
saddle points, i.e., binning in the $x$ coordinate. 

\begin{figure*}
\centering
\includegraphics[width=0.9\textwidth]{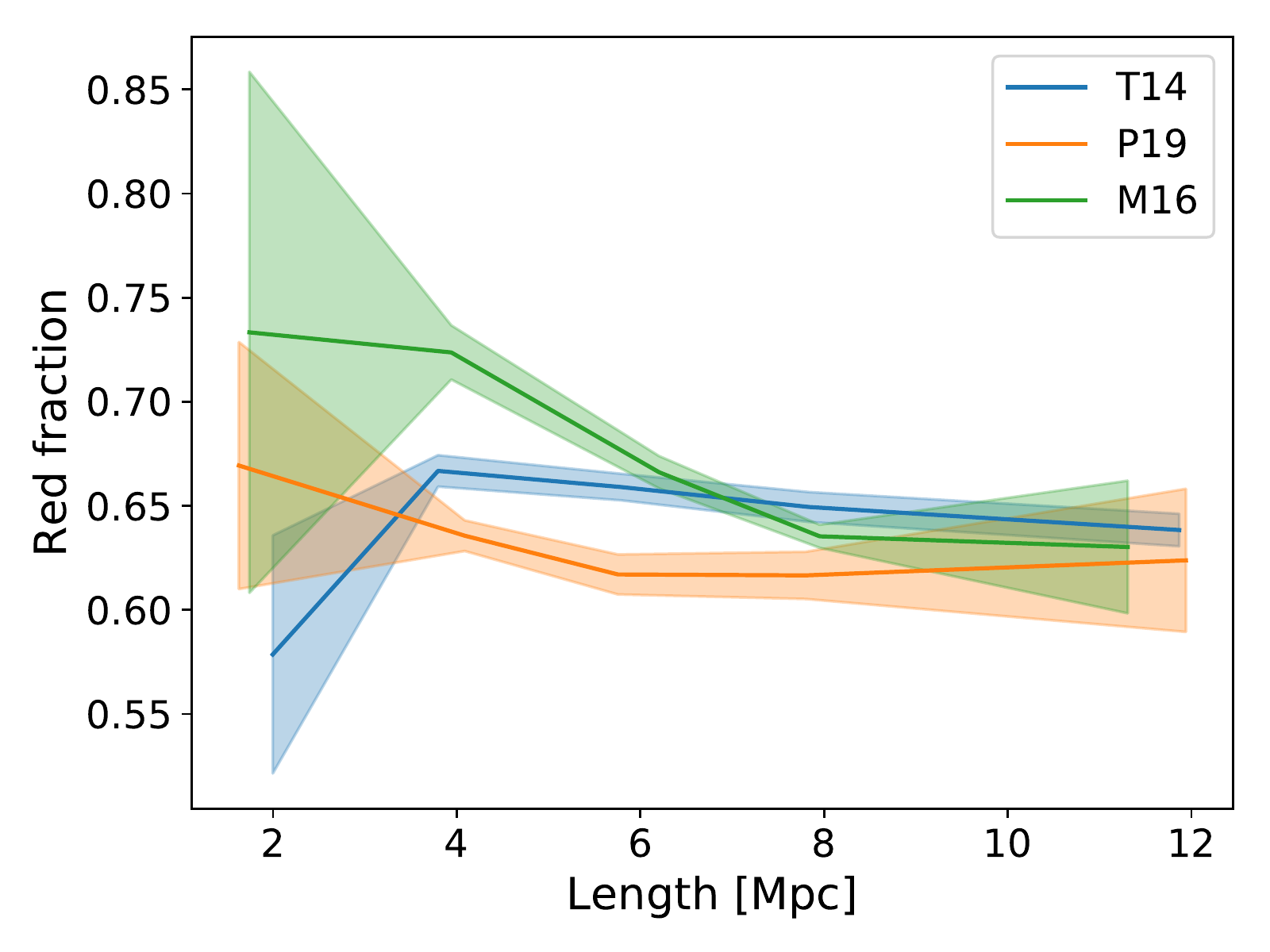}
\caption[Red fractions]
{Fraction of red galaxies in filaments for all the catalogues.
It is possible to notice a dependence with the filament length, longer
filaments have less fraction of red galaxies.}
\label{fraccion_rojas}
\end{figure*}

In Fig. \ref{fraccion_rojas}, we show the dependence of the red fraction galaxies in filaments with the 
filament's length. We see a general trend that larger filaments have less number of red fraction, tending towards 
the mean value.  That has sense because of the overdensity values of the filaments increases for smaller filaments. In Temple's case there is a lot of dispersion, and for the stochasticity of their method not bit trend is found.

\subsection{Colour vs. Luminosity}
\label{col_vs_mag}
Galaxy properties are not independent of each other, and they are affected by environment. 
As we move from lower to higher density, galaxies are more likely to be brighter, redder, to have
earlier morphology, and lower star formation rate. Regarding the broad-band photometrical 
properties of
galaxies, it has been shown by \citet{blanton} that colour is the property that correlates best with
local density. In systems of galaxies, colour is also the best predictor of both: the system mass,
and the distance to the centre of the system \citep{HM2:2006,MCM:2008}. We have shown in the 
previous subsection how colour correlates with the distance to the filament's axis, having red 
galaxies a higher overdensity profile compared to blue ones. In this subsection we explore how
absolute magnitude is related to the distance to filament's axis, and compare the resulting
overdensity profiles with those in the previous subsection, addressing the question of which 
property is a better predictor of distance.

\begin{figure}
\centering
\includegraphics[width=0.5\textwidth]{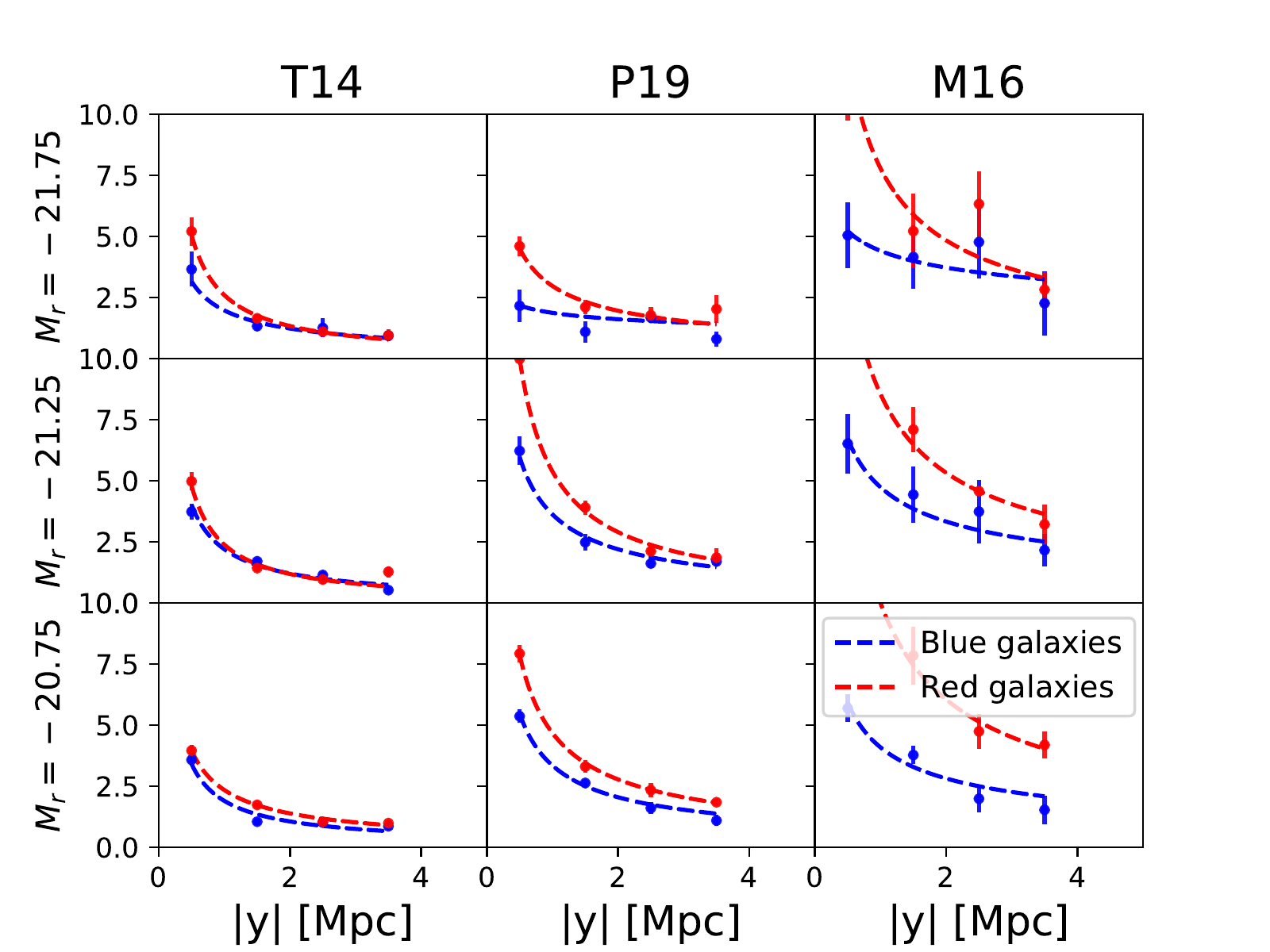}
\caption[Stacking $4 {\rm Mpc}$]
{
The overdensity profile of red ($ (g - r) > 0.7$) and blue ($(g - r) < 0.7$) galaxies, of different magnitudes. From the top to the bottom: Galaxies of
$M_{r} = -21.75 \pm 0.25$, $M_{r} = -21.25 \pm 0.25$ and $M_{r} = -20.75 \pm 0.25$.}
\label{f17} 
\end{figure}

We focus our study in our samples of filaments with lengths in the range 
$3-5~h^{-1}{\rm Mpc}$, and consider three absolute magnitude bins, centred in 
$M_r-5\log(h)=-20.75, -21.25$, and $-21.75$, and width 0.5.
We then compute the overdensity profiles of blue and red galaxies with absolute magnitudes
within these ranges. Resulting profiles are show in Fig. \ref{f17}.
Since our overdensity profiles are a particular type of two-point correlation functions,
we fit a power-law of the form $I(|y|)=A|y|^{-\gamma}$, to all profiles in 
Fig. \ref{f17}, which we show as dashed lines. Resulting amplitudes and
power-law indexes are quoted in Table \ref{tab:correlacion}.
We find that power-law is a good fit in most cases considered, as can be seen from Fig. \ref{f17}.
Both parameters show, in general, a larger variation from blue to red galaxies at fixed absolute magnitude,
than as a function of luminosity at a fixed colour type. 
For both, blue and red galaxies, and at fixed absolute 
magnitude, amplitudes are in agreement with the results discussed in the previous subsection. 
Sorting according to increasing values of the amplitude we have the filaments by T14, P19, and M16.
There is no such a clear ranking when we compare power-law indexes, but in general it would be 
M16, P19 and T14, for increasing $\gamma$, which is in qualitative agreement with Figs. $10-14$. 
A singular systematic feature of P19 filaments is that both, the largest amplitude, and the
largest power-law index, are obtained for galaxies with $M_r-5\log(h)\sim-21.25$ which are the 
galaxies that P19 use to define their filaments.

Similar results are obtained for longer filaments. We do not include them here 
in order to not extend further the paper.
The main conclusion of this subsection is that the overdensity around filaments is better
traced by colour than luminosity.
We recall again that in this paper we are probing the bright
end of the galaxy population, thus our results concern only to these bright galaxies.

\begin{table*}
    \centering
    \begin{tabular}{lccccccc}
    & & \multicolumn{2}{c}{T14}  & \multicolumn{2}{c}{M16}  & \multicolumn{2}{c}{P19}  \\
    \hline
    Colour & $M_r$ & $A$ & $\gamma$ & $A$ & $\gamma$ & $A$ & $\gamma$ \\
    \hline
    Blue & $-20.75\pm0.25$ & 
           $1.90\pm0.04$ & $0.84\pm0.02$ & 
           $4.1\pm0.1$ & $0.54\pm0.02$ & 
           $3.35\pm0.02$ & $0.71\pm0.01$ \\
         & $-21.25\pm0.25$ & 
           $2.20\pm0.06$ & $0.87\pm0.02$ & 
           $4.7\pm0.2$ & $0.51\pm0.01$ &
           $3.6\pm0.1$ & $0.72\pm0.01$ \\
         & $-21.75\pm0.25$ & 
           $1.97\pm0.05$ & $0.68\pm0.02$ & 
           $4.4\pm0.4$ & $0.24\pm0.03$ & 
           $1.9\pm0.3$ & $0.2\pm0.1$ \\
    \hline
    Red  & $-20.75\pm0.25$ & 
           $2.34\pm0.01$ & $0.76\pm0.01$ & 
           $10.0\pm0.2$ & $0.72\pm0.01$ & 
           $4.68\pm0.01$ & $0.75\pm0.01$ \\
         & $-21.25\pm0.25$ & 
           $2.38\pm0.07$ & $1.01\pm0.03$ & 
           $8.6\pm0.2$ & $0.68\pm0.01$ &
           $5.38\pm0.02$ & $0.89\pm0.01$ \\
         & $-21.75\pm0.25$ & 
           $2.59\pm0.01$ & $0.96\pm0.01$ & 
           $8\pm2$ & $0.69\pm0.04$ & 
           $2.97\pm0.04$ & $0.60\pm0.01$ \\
    \hline
    \end{tabular}
    \caption{Best fit power-law parameters to the galaxy overdensity profiles of Fig. \ref{f17}.}
    \label{tab:correlacion}
\end{table*}

\section{Conclusions} 
\label{sec:conclusions}

In this paper we present a comparison between different catalogues of cosmological filaments
identified by different methods: \citet{pereyra_submitted}, \citet{Tempel2014} and \citet{martinez}.
They differ notably in the way filaments are defined. T14's algorithm is
different than the others in the sense that it uses geometrical assumptions,
while the remaining algorithms take in account physical information from the
galaxies. 
This leads to significant differences between those kinds of
algorithms that are reflected in the properties of the catalogues.

It is important to note that since these algorithms do not detect walls, some
detected filaments could be rather part of walls than a filament itself. This
has to be taken into account because walls are different objects and there are
other phenomena occurring in them. 

The different algorithms do not find the same filaments, instead they find them
in common dense regions and in different amounts. Their properties such as the
distributions of length, elongation, and redshift vary in each catalogue,
T14 filaments are longer and at lower redshift in general in
comparison with P19 and M16 catalogues, on the other hand P19 finds sets of filaments more irregular shaped. 

Other quantities defined in
this work such as the relations length and overdensity, luminosity, average
luminosity, etc. also differ between catalogues, T14 filaments are less
over-dense than the other catalogues, and their average luminosity per galaxy
is indistinguishable from a random set of galaxies. On the other hand,
P19 and M16 are more over-dense and the average luminosity of the
galaxies that belong to them are higher than what it would be if compared with
a random galaxy catalogue. There is a correlation between the filaments's
length and overdensity, the overdensity decreases with long filaments, which
suggests that short filaments are `stronger' than long filaments in agreement with
\citet{bond}. In the case of T14 there could be an over-estimation of the width for long filaments that would cover uncorrelated volumes with these objects.

Through the analysis of the bright-end of the galaxy luminosity functions 
in different environments (groups, filaments and field galaxies), we find that galaxies
in filaments have characteristic magnitude intermediate between the field and group counterparts. 
The most interesting feature is the $\alpha=-0.26$ value of P19 filaments that, given the
fact that we use galaxies brighter than $M_r=-20.5$, is indicating a more convex shape of P19
filaments bright-end luminosity function. 
The luminosity function does not vary much with filament length, but there is a tendency of shorter
filaments to have brighter characteristic magnitude. Overall, the luminosity functions of T14 and M16
filaments are consistent within errors.

We also develop an statistical tool based on a stacking method that allows us
to investigate the spatial distribution of galaxies in and around filaments.
With this method we show that the filaments from different catalogues, constructed with various
methods, when stacked, they look different, exhibiting diverse features. The
one-dimensional overdensity profiles of galaxies also differ, the T14 catalogue shows a steeper
density distribution, while P19 and M16 are similar to each other
and more extended. The red and blue galaxy distributions also differ between
catalogues, the ones that use physical information 
show more variation between both red and blue profiles while with the T14 catalogue the difference is
lesser. 
 We also explore the dependence of the overdensity profiles on luminosity. While there are trends with luminosity, they are in general weaker compared to the dependences on colour.
Filament length is an important factor: shorter filaments show a higher overdensity of galaxies.
The fraction of red galaxies also vary with
the filament length for the M16 and P19 catalogues, such
dependence is not found with T14. The red fraction dependence with the
length and the position along the filament's axis has been explored too, but
the results that we find are too noisy to reach clear conclusions.

\section*{Acknowledgements}
We thank the anonymous referee for  improving the paper, in particular, the section \ref{col_vs_mag} resulted from exploring in depth her/his comments.
We thank Ana Laura O'Mill, Andr\'es Ruiz, Manuel Merch\'an, Mario Sgr\'o,
and Dante Paz for useful comments and discussions.
In particular Dr. O'Mill for providing us the tracers galaxy data sample.
This work was partially supported by Consejo Nacional de Investigaciones 
Cient\'{\i}ficas y T\'ecnicas (CONICET) grants  (PIP 11220130100365CO), Argentina, FONCyT PICT-2016-4174
and Secretar\'{\i}a de Ciencia y Tecnolog\'{\i}a de la Universidad Nacional 
de C\'ordoba (SeCyT-UNC, Argentina).

\bibliographystyle{mnras}
\bibliography{biblio}

\end{document}